\pdfoutput=1

\documentclass[conference]{acmsiggraph}

\usepackage{booktabs}
\usepackage{microtype}
\usepackage{url}
\usepackage{color}

\definecolor{darkgreen}{rgb} {0.0,0.5,0.0}
\definecolor{darkblue}{rgb}  {0.0,0.0,0.7}
\definecolor{darkred}{rgb}   {0.7,0.0,0.0}
\definecolor{brown}{rgb}     {0.5,0.3,0.2}
\definecolor{lightgray}{rgb} {0.9,0.9,0.9}
\definecolor{lightred}{rgb}  {0.7,0.2,0.2}

\usepackage{pifont}
\usepackage[normalem]{ulem}
\usepackage{booktabs}
\usepackage{multirow}
\usepackage{subcaption}
\usepackage{ifthen}
\usepackage{ifpdf}
\usepackage{amsmath}
\usepackage[pdftex,dvipsnames]{xcolor}
\usepackage[prependcaption]{todonotes}
\usepackage[T1]{fontenc}
\usepackage[scaled=0.85]{beramono}

\ifpdf \usepackage{graphicx} \pdfcompresslevel=9
\else \usepackage[dvips]{graphicx} \fi
\usepackage{graphicx}
\ifpdf
  \usepackage{epstopdf}
\fi

\newcommand{\piko}{Piko}
\newcommand{\Piko}{\piko}
\newcommand{\pikoc}{\texttt{Pikoc}}
\newcommand{\schedule} {\texttt{Schedule}}
\newcommand{\assignbin}{\texttt{AssignBin}}
\newcommand{\process}  {\texttt{Process}}
\newcommand{\kw}[1]{\texttt{#1}}
\newcommand{\stage}[1]{\texttt{#1}}

\usepackage{algorithm}
\usepackage{algpseudocode}

\usepackage{parskip}
\usepackage[labelfont=bf,textfont=it]{caption}

    \hypersetup{pdftitle={piko}}
    \hypersetup{colorlinks=true}
    \hypersetup{urlcolor=black}
    \hypersetup{citecolor=blue}
    \hypersetup{linkcolor=black}
    \hypersetup{breaklinks=true}
    \hypersetup{pdfstartview=FitV}
    \hypersetup{pdfstartpage=1}
    \hypersetup{pdfdisplaydoctitle=true}

\usepackage{color}
\definecolor{gray}{rgb}{0.4,0.4,0.4}
\definecolor{dg}{rgb}{0.0,0.65,0.0}
\definecolor{db}{rgb}{0.0,0.0,0.8}
\definecolor{dp}{rgb}{0.8,0.0,0.8}
\definecolor{lb}{rgb}{0.0,0.5,0.6}
\usepackage{listings}
\presetkeys{todonotes}{fancyline, inline, color=red!60}{}

\newcommand{\stanley}[1]{\todo[inline,author=Stanley,color=blue!20]{#1}}
\newcommand{\anjul}  [1]{\todo[inline,author=Anjul,color=green!20]{#1}}
\newcommand{\john}   [1]{\todo[inline,author=John,color=red!20]{#1}}
\newcommand{\kerry}  [1]{\todo[inline,author=Kerry,color=brown!30]{#1}}
\newcommand{\tim}[1]{\todo[inline,author=Tim's Opinion,color=red!60]{#1}}

\newcommand{\nothing}[1]{}

\newcommand{\final}{1}

\ifthenelse{\equal{\final}{1}}
{
\renewcommand{\stanley}[1]{}
\renewcommand{\anjul}[1]{}
\renewcommand{\john}[1]{}
\renewcommand{\todo}[1]{}
\renewcommand{\kerry}[1]{}

\renewcommand{\tim}[1]{}
}
{}
\newcommand{\secref} [1]{Section~\ref{sec:#1}}
\newcommand{\figref} [1]{Figure~\ref{fig:#1}}
\newcommand{\tabref} [1]{Table~\ref{tab:#1}}

\newcommand{\listref}[1]{Listing~\ref{list:#1}}

\TOGonlineid{0414}

\TOGvolume{0}
\TOGnumber{0}
\TOGarticleDOI{1111111.2222222}

\TOGprojectURL{}
\TOGvideoURL{}
\TOGdataURL{}
\TOGcodeURL{}

\title{\piko: A Design Framework for Programmable Graphics Pipelines}

\author{  Anjul Patney\thanks{email:apatney@nvidia.com} \and
          Stanley Tzeng\thanks{email:stzeng@nvidia.com}   \and
          Kerry A. Seitz, Jr.\thanks{email:kaseitz@ucdavis.edu}  \and
         John D.\ Owens\thanks{email:jowens@ece.ucdavis.edu}
        \affiliation{University of California, Davis} 
}
\pdfauthor{Anjul Patney, Stanley Tzeng, Kerry A. Seitz Jr., John D. Owens}

\keywords{graphics pipelines, parallel computing}

\begin{document}

\teaser{
  \includegraphics[page=2,width=0.7\textwidth,clip,trim=0.0in 1.9in 0.0in 0.0in]
  {images/system/teasers.pdf}

  \caption{\piko{} is a framework for designing and implementing programmable
  graphics pipelines that can be easily retargeted to different
  application configurations and architectural targets. \piko{}'s input
  is a functional and structural description of the desired graphics
  pipeline, augmented with a per-stage grouping of computation into
  spatial bins (or tiles), and a scheduling preference for these bins.
  Our compiler generates efficient implementations of the input pipeline
  for multiple architectures and allows the programmer to tweak these
  implementations using simple changes in the bin configurations and scheduling
  preferences.
  \label{fig:teaser}
  \label{fig:system-overview}
  }
}


\maketitle


\begin{abstract}

  We present \piko{}, a framework for designing, optimizing, and
  retargeting implementations of graphics pipelines on multiple
  architectures. \piko{} programmers express a graphics pipeline by
  organizing the computation within each stage into spatial bins and
  specifying a scheduling preference for these bins. Our compiler, \pikoc{},
  compiles this input into an optimized implementation targeted to a
  massively-parallel GPU or a multicore CPU\@.

  \piko{} manages work granularity in a programmable and flexible
  manner, allowing programmers to build load-balanced parallel
  pipeline implementations, to exploit spatial and producer-consumer
  locality in a pipeline implementation, and to explore tradeoffs
  between these considerations. We demonstrate that \piko{} can
  implement a wide range of pipelines, including rasterization, Reyes,
  ray tracing, rasterization/ray tracing hybrid, and deferred
  rendering. \piko{} allows us to implement efficient graphics
  pipelines with relative ease and to quickly explore design
  alternatives by modifying the spatial binning configurations and
  scheduling preferences for individual stages, all while delivering
  real-time performance that is within a factor six of state-of-the-art
  rendering systems.

\end{abstract}


\begin{CRcatlist}
  \CRcat{I.3.1}{Computer Graphics}{Hardware architecture}{Parallel processing};
  \CRcat{I.3.2}{Computer Graphics}{Graphics Systems}{Stand-alone systems}
\end{CRcatlist}


\keywordlist


\TOGlinkslist


\copyrightspace


\section{Introduction}

Renderers in computer graphics often build upon an underlying graphics pipeline:
a series of computational stages that transform a scene description into an
output image. Conceptually, graphics pipelines can be represented as a graph with
stages as nodes, and the flow of data along directed edges of the graph.
While some renderers target the special-purpose hardware pipelines built into
graphics processing units (GPUs), such as the OpenGL/Direct3D
pipeline (the ``OGL/D3D pipeline''), others use pipelines implemented in software,
either on CPUs or, more recently, using the programmable capabilities of
modern GPUs. This paper concentrates on the problem of implementing a
graphics pipeline that is both highly programmable and high-performance by
targeting programmable parallel processors like GPUs.

Hardware implementations of the OGL/D3D pipeline are extremely efficient, and
expose programmability through shaders which customize the behavior of
stages within the pipeline. However, developers cannot easily customize the
structure of the pipeline itself, or the function of non-programmable stages.
This limited programmability makes
it challenging to use hardware pipelines to implement other types
of graphics pipelines, like ray tracing, micropolygon-based pipelines,
voxel rendering, volume rendering, and hybrids that
incorporate components of multiple pipelines. Instead, developers have
recently begun using programmable GPUs to implement these pipelines in
software (Section~\ref{sec:programmable-graphics-abstractions}),
allowing their use in interactive applications.

Efficient implementations of graphics pipelines are complex: they must
consider parallelism,
load balancing, and locality within the bounds of a restrictive
programming model. In general, successful pipeline implementations
have been narrowly customized to a particular pipeline and
often to a specific hardware target. The abstractions and
techniques developed for their implementation are not easily
extensible to the more general problem of creating efficient yet
structurally- as well as functionally-customizable, or programmable pipelines. Alternatively, researchers have explored more general
systems for creating programmable pipelines, but these systems compare
poorly in performance against more customized pipelines, primarily
because they do not exploit specific characteristics of the pipeline
that are necessary for high performance.

Our framework, \piko{}, builds on spatial bins, or tiles,
to expose an interface which allows pipeline implementations to exploit
load-balanced parallelism and both producer-consumer and spatial
locality, while still allowing high-level programmability. Like traditional
pipelines, a \piko{} pipeline consists of a series of stages
(Figure~\ref{fig:system-overview}), but we further decompose those
stages into three abstract \emph{phases} (Table~\ref{tab:phases}).
These phases expose the salient characteristics of the pipeline that
are helpful for achieving high performance. \piko{}
pipelines are compiled into efficient software implementations for multiple target
architectures using our compiler, \pikoc{}. \pikoc{} uses the
LLVM framework~\cite{Lattner:2004:LLVM} to automatically translate
user pipelines into the LLVM intermediate representation (IR) before
converting it into code for a target architecture.

We see two major differences from previous work. First, we describe an
abstraction and system for designing and implementing generalized
programmable pipelines rather than targeting a single programmable
pipeline. Second, our abstraction and implementation incorporate
spatial binning as a fundamental component, which we demonstrate is a
key ingredient of high-performance programmable graphics pipelines.

The key contributions of this paper include:

\begin{itemize}
\item Leveraging \emph{programmable binning for spatial locality} in
  our abstraction and implementation, which we demonstrate is critical
  for high performance;
\item Factoring pipeline stages into 3 phases, \assignbin, \schedule,
  and \process, which allows us to flexibly exploit spatial locality and
  which enhances portability by factoring stages into
  architecture-specific and -independent components;
\item Automatically identifying and exploiting opportunities for
  compiler optimizations directly from our pipeline descriptions; and
\item A compiler at the core of our programming system that
  automatically and effectively generates pipeline code from the
  \piko{} abstraction, achieving our goal of constructing
  easily-modifiable and -retargetable, high-performance, programmable
  graphics pipelines.
\end{itemize}

\section{Programmable Graphics Abstractions} 
\label{sec:related_work}
\label{sec:programmable-graphics-abstractions}

Historically, graphics pipeline designers have attained flexibility through the
use of programmable shading. Beginning with a fixed-function pipeline with
configurable parameters, user programmability began in the form of register
combiners, expanded to programmable vertex and fragment shaders (e.g.,
Cg~\cite{Mark:2003:CAS}), and today encompasses tessellation, geometry, and even
generalized compute shaders in Direct3D~11. Recent research has also proposed
programmable hardware stages beyond shading, including a delay stream between
the vertex and pixel processing units~\cite{Aila:2003:DSF} and the programmable
culling unit~\cite{Hasselgren:2007:PTP}.

The rise in programmability has led to a number of innovations beyond the OGL/D3D pipeline.
Techniques like deferred
rendering (including variants like tiled-deferred lighting in compute shaders,
as well as subsequent approaches like ``Forward+'' and clustered forward rendering), amount
to building alternative pipelines that schedule work differently and exploit different
trade-offs in locality, parallelism, and so on.
In fact, many modern games already
implement a deferred version of forward rendering to reduce the cost
of shading and reduce the number of rendering
passes~\cite{Andersson:2009:PGI}.

Recent research uses the programmable aspects of modern GPUs to
implement entire pipelines in software. These efforts include
RenderAnts, which implements a GPU Reyes
renderer~\cite{Zhou:2009:RIR}; cudaraster~\cite{Laine:2011:HSR}, which
explores software rasterization on GPUs; VoxelPipe, which targets
real-time GPU voxelization~\cite{Pantaleoni:2011:VAP}, and the
Micropolis Reyes renderer~\cite{Weber:2015:PRA}. The
popularity of such explorations demonstrates that entirely
programmable pipelines are not only feasible but desirable as well.
These projects, however, target a single specific pipeline for one
specific architecture, and as a consequence their implementations
offer limited opportunities for flexibility and reuse.

A third class of
recent research seeks to rethink the historical approach to
programmability, and is hence most closely related to our work.
GRAMPS~\cite{Sugerman:2009:GAP} introduces a programming model that
provides a general set of abstractions for building parallel graphics
(and other) applications. Sanchez et al.~\shortcite{Sanchez:2011:DFG}
implemented a multi-core x86 version of GRAMPS\@. NVIDIA's high-performance
programmable ray tracer OptiX~\cite{Parker:2010:OAG} also allows arbitrary
pipes, albeit with a custom scheduler specifically designed for their GPUs.
By and large, GRAMPS addresses expression
and scheduling at the level of pipeline organization, but does not
focus on handling efficiency concerns within individual stages.
Instead, GRAMPS successfully focuses on programmability,
heterogeneity, and load balancing, and relies on the efficient design
of inter-stage sequential queues to exploit producer-consumer
locality. The latter is in itself a challenging implementation task
that is not addressed by the GRAMPS abstraction. The principal
difference in our work is that instead of using queues, we use 2D
tiling to group computation in a manner that helps balance parallelism
with locality and is more optimized towards graphcal workloads. While
GRAMPS proposes queue sets to possibly expose
parallelism within a stage (which may potentially support spatial
bins), it does not allow any flexibility in the scheduling strategies
for individual bins, which, as we will demonstrate, is important to
ensure efficiency by tweaking the balance between spatial/temporal
locality and load balance. \piko{} also merges user stages together into
a single kernel for efficiency purposes.  GRAMPS relies directly on the
programmer's decomposition of work into stages so that fusion,
which might be a target-specific optimization, must be done at
the level of the input pipeline specification.

Peercy et al.~\shortcite{Peercy:2000:IMP} and
FreePipe~\cite{Liu:2010:FAP} implement an entire OGL/D3D pipeline in
software on a GPU, then explore modifications to their pipeline to
allow multi-fragment effects. These GPGPU software rendering pipelines
are important design points; they describe and analyze optimized
GPU-based software implementations of an OGL/D3D pipeline, and are thus
important comparison points for our work. We demonstrate that our
abstraction allows us to identify and exploit optimization
opportunities beyond the FreePipe implementation.

Halide~\cite{Ragan-Kelley:2012:DAS} is a domain-specific embedded
language that permits succinct, high-performance implementations of
state-of-the-art image-processing pipelines. In a manner similar to
Halide, we hope to map a high-level pipeline description to a
low-level efficient implementation. However, we employ this strategy
in a different application domain, programmable graphics, where data
granularity varies much more throughout the pipeline and dataflow is
both more dynamically varying and irregular.
Spark~\cite{Foley:2011:SMC} extends the flexibility of shaders such
that instead of being restricted to a single pipeline stage, they can
influence several stages across the pipeline. Spark allows such
shaders without compromising modularity or having a significant impact
on performance, and in fact Spark could be used as a shading language
to layer over pipelines created by \piko. We share design goals that
include both flexibility and competitive performance in the same spirit
as Sequoia~\cite{Fatahalian:2006:SPT} and StreamIt~\cite{Thies:2002:SAL}
 in hopes of abstracting out
the computation from the underlying hardware.

\section{Spatial Binning}

Both classical and modern graphics systems often render images by dividing the
screen into a set of regions, called tiles or spatial \emph{bins}, and processing
those bins in parallel. Examples include tiled rasterization, texture and
framebuffer memory layouts, and hierarchical depth buffers. Exploiting spatial
locality through binning has five major advantages. First, it prunes away
unnecessary work associated with the bin---primitives not affecting a bin are
never processed. Second, it allows the hardware to take advantage of data and
execution locality within the bin itself while processing (for example, tiled
rasterization leads to better locality in a texture cache). Third,
many pipeline stages may have a natural granularity of work that is
most efficient for that particular stage; binning allows programmers
to achieve this granularity at each stage by tailoring the size of bins.
Fourth, it exposes an additional level of data
parallelism, the parallelism between bins. And fifth, grouping computation into
bins uncovers additional opportunities for exploiting producer-consumer locality
by narrowing working-set sizes to the size of a bin. 

\begin{table}[t]
\centering
\begin{small}
\begin{tabular}{p{1.1in}l}
\toprule
\textbf{Reference-Image Binning}    & PixelFlow~\cite{Olano:1998:ASL} \\
                                    & Chromium~\cite{Humphreys:2002:CAS} \\ \midrule
\textbf{Interleaved Rasterization}  & AT\&T Pixel Machine~\cite{Potmesil:1989:TPM} \\
                                    & SGI InfiniteReality~\cite{Montrym:1997:IAR} \\ 
                                    & NVIDIA Fermi~\cite{Purcell:2010:FTR} \\ \midrule
\textbf{Tiled Rasterization/}       & RenderMan~\cite{Apodaca:1990:RPT} \\
\textbf{Chunking}                   & cudaraster~\cite{Laine:2011:HSR} \\
                                    & ARM Mali~\cite{Olson:2012:STP} \\
                                    & PowerVR~\cite{Imagination:2011:PSG} \\
                                    & RenderAnts~\cite{Zhou:2009:RIR} \\ \midrule
\textbf{Tiled Depth-Based}          & Lightning-2~\cite{Stoll:2001:LAH} \\
\textbf{Composition}                &  \\ \midrule
\textbf{Bin Everywhere}             & Pomegranate~\cite{Eldridge:2000:PAF} \\ \bottomrule
\end{tabular}

\end{small}

\caption{Examples of Binning in Graphics Architectures. We characterize pipelines 
based on when spatial binning occurs. Pipelines that bin prior to the geometry stage 
are classified under `reference-image binning'. Interleaved and tiled
rasterization pipelines typically bin between the geometry and rasterization
stage. Tiled depth-based composition pipelines bin at the sample or composition stage. 
Finally, `bin everywhere' pipelines bin after every stage by re-distributing the primitives in dynamically updated queues.}

\label{tab:binningtable}
\end{table}

\nothing{
\textbf{Reference-Image Binning} & PixelFlow~\cite{Olano:1998:ASL} \\
& Princeton Wall~\cite{Li:2000:BAU} \\
& WarpEngine~\cite{Popescu:2000:TWA} \\
& WireGL~\cite{Humphreys:2001:WAS} \\
& Sage~\cite{Deering:2002:TSG} \\
& Chromium~\cite{Humphreys:2002:CAS} \\
& D3DPR~\cite{Liu:2005:DBL} \\ \midrule
\textbf{Interleaved Rasterization} & AT\&T Pixel Machine~\cite{Potmesil:1989:TPM} \\
& SGI RealityEngine~\cite{Akeley:1993:RG} \\
& SGI InfiniteReality~\cite{Montrym:1997:IAR} \\ \midrule
\textbf{Tiled Rasterization/} & Pixel-Planes 5~\cite{Fuchs:1989:P5A} \\
\textbf{Chunking} & RenderMan~\cite{Apodaca:1990:RPT} \\
& Talisman~\cite{Torborg:1996:TCR} \\
& ARM Mali~\cite{Stevens:2006:ARM} \\
& Imagination's PowerVR MBX \\& \hspace{1em}\cite{PowerVR:2008:PVR} \\
& Larrabee~\cite{Seiler:2008:LAM} \\
& RenderAnts~\cite{Zhou:2009:RIR} \\
& Lazy Object-Space Shading~\cite{Burns:2010:LOS} \\
& FreePipe~\cite{Liu:2010:FAP} \\
& CudaRaster~\cite{Laine:2011:HSR} \\ \midrule
\textbf{Tiled Depth-Based}  & Sepia2~\cite{Heirich:1999:SDV} \\
\textbf{Composition} & Lightning-2~\cite{Stoll:2001:LAH} \\
& RenderAnts~\cite{Zhou:2009:RIR} \\\midrule
\textbf{Bin Everywhere} & Pomegranate~\cite{Eldridge:2000:PAF} \\ \bottomrule }

Spatial binning has been a key part of graphics systems dating to some
of the earliest systems. The Reyes pipeline~\cite{Cook:1987:TRI} tiles
the screen, rendering one bin at a time to avoid working sets that are
too large; Pixel-Planes~5~\cite{Fuchs:1989:P5A} uses spatial binning
primarily for increasing parallelism in triangle rendering and other
pipelines. More recently, most major GPUs use some form of spatial
binning, particularly in
rasterization~\cite{Olson:2012:STP,Purcell:2010:FTR}.

Recent software renderers written for CPUs and GPUs
also make extensive use of screen-space tiling:
RenderAnts~\cite{Zhou:2009:RIR} uses buckets to limit memory usage
during subdivision and sample stages, cudaraster~\cite{Laine:2011:HSR}
uses a bin hierarchy to eliminate redundant work and provide more
parallelism, and VoxelPipe~\cite{Pantaleoni:2011:VAP} uses
tiles for both bucketing purposes and exploiting spatial locality.
Table~\ref{tab:binningtable} shows examples of graphics systems that
have used a variety of spatial binning strategies.

The advantages of spatial binning are so compelling that we believe, and will
show, that exploiting spatial binning is a crucial component for performance in
efficient implementations of graphics pipelines.  Previous work in
software-based pipelines that take advantage of binning has focused on specific,
hardwired binning choices that are narrowly tailored to one particular pipeline.
In contrast, the \piko{} abstraction encourages pipeline designers to express
pipelines and their spatial locality in a more general, flexible,
straightforward way that exposes opportunities for binning optimizations and
performance gains.

\section{Expressing Pipelines Using \piko}
\label{sec:abstraction}

\subsection{High-Level Pipeline Abstraction}

Graphics algorithms and APIs are typically described as pipelines
(directed graphs) of simple stages that compose to create complex
behaviors. The OGL/D3D abstraction is described in this fashion, as
are Reyes and GRAMPS, for instance. Pipelines aid understanding, make
dataflow explicit, expose locality, and permit reuse of individual
stages across different pipelines. At a high level, the \piko{}
pipeline abstraction is identical, expressing computation within
stages and dataflow as communication between stages. \piko{} supports
complex dataflow patterns, including a single stage feeding input to
multiple stages, multiple stages feeding input to a single stage, and
cycles (such as Reyes recursive splitting).

\begin{figure*}[t]
\begin{center}
  \includegraphics[width=0.65\textwidth,clip,trim=0.0in 0.0in 0.0in 0.0in]
  {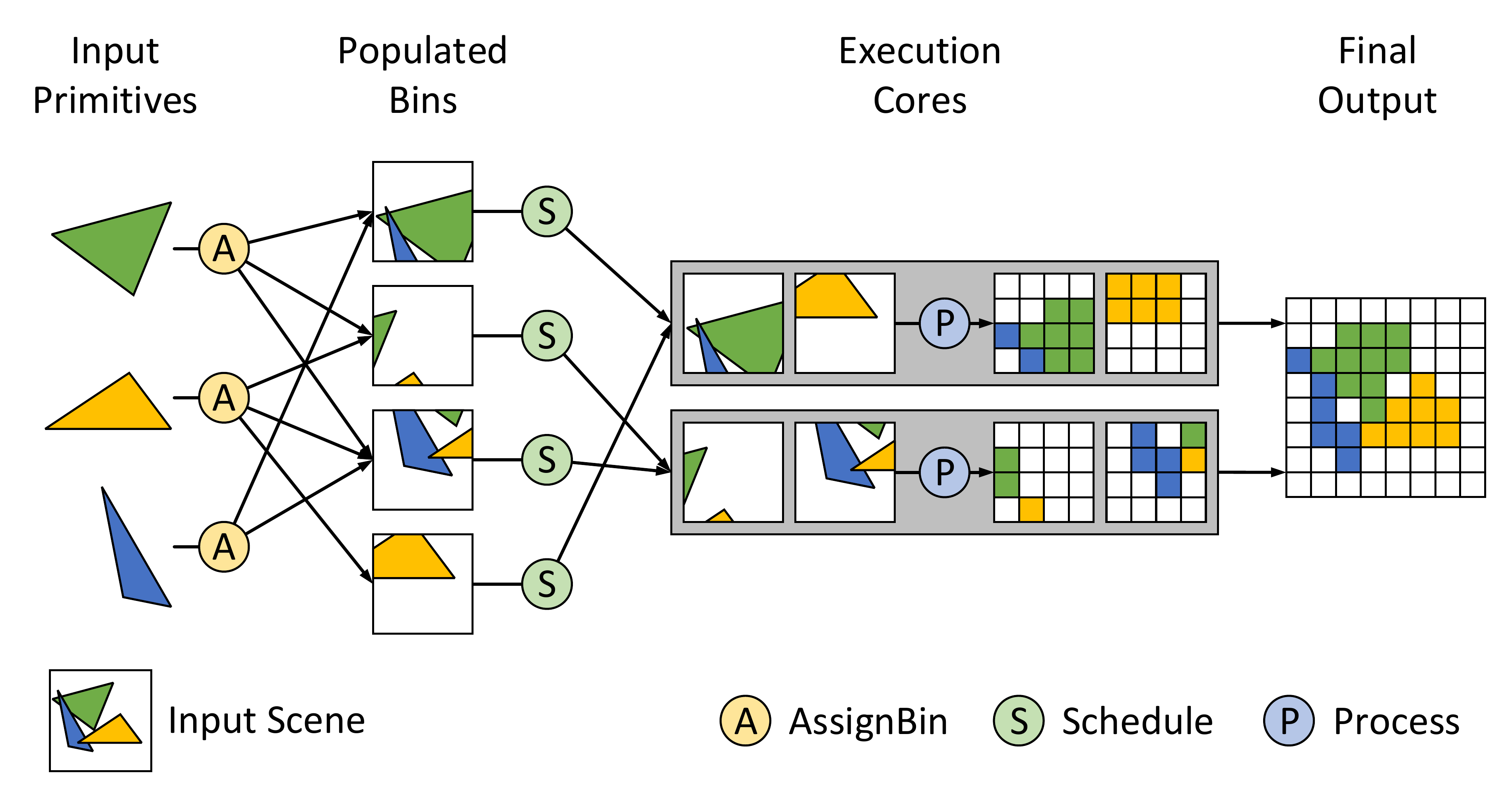}

  \caption[Phases in a \piko{} stage] 
  {The three phases of a \piko{} stage. This diagram shows the role of
  \assignbin{}, \schedule{}, and \process{} in the scan conversion of a list of
  triangles using two execution cores. Spatial binning helps in (a) extracting 
  load-balanced parallelism by assigning triangles into smaller, more uniform, spatial
  bins, and (b) preserving spatial locality within each bin by grouping together 
  spatially-local data. The three phases help fine-tune how computation is grouped and
  scheduled, and this helps quickly explore the optimization space of an implementation.}



\label{fig:dataflow}

\end{center}
\end{figure*}

Where the abstraction differs is within a pipeline stage.
Consider a \textsc{Baseline} system that would implement one of the above pipelines as
a set of separate per-stage
kernels, each of which distributes its work to available parallel cores, and
the implementation connects the output of one stage to the input of the next through
off-chip memory. Each instance of a \textsc{Baseline} kernel would run over the
entire scene's intermediate data, reading its input from off-chip memory and writing
its output back to off-chip memory. This implementation would have ordered
semantics and distribute work in each stage in FIFO order.

Our \textsc{Baseline} would end up making poor use of both the producer-consumer
locality between stages and the spatial locality within and between
stages. It would also require a rewrite of each stage to target a different
hardware architecture.
\piko{} specifically addresses these issues by balancing between enabling
productivity and
portability through a high-level programming model, while specifying
enough information to allow high-performance implementations. The distinctive
feature of the abstraction is the ability to cleanly separate the implementation
of a high-performance graphics pipeline into
separable, composable concerns, which provides two main benefits:

\begin{itemize}
\item It facilitates modularity and architecture independence.
\item It integrates locality and spatial binning in a way that
  exposes opportunities to explore the space of optimizations involving
  locality and load-balance.
\end{itemize}

\begin{table}[t]
\begin{center}

\begin{tabular}{lll}
\toprule
\textbf{Phase}            & \textbf{Granularity}     & \textbf{Purpose}\\
\midrule
\assignbin                 & Per-Primitive            & How to group computation?\\
\midrule
\multirow{2}{*}{\schedule} & \multirow{2}{*}{Per-Bin} & When to compute? \\
                           &                          & Where to compute?\\
\midrule
\process                   & Per-Bin or   & How to compute?\\
                          & Per-primitive & \\
\bottomrule

\end{tabular}

\caption{Purpose and granularity for each of the three phases during each stage.
We design these phases to cleanly separate the key steps in a
pipeline built around spatial binning. Note: we consider \process{} a per-bin
operation, even though it often operates on a per-primitive basis.}

\label{tab:phases}
\end{center}
\end{table}

For the rest of the paper, we will use the following terminology
for the parts of a graphics pipeline. Our pipelines are expressed as
directed graphs where each node represents a self-contained functional
unit or a \emph{stage}. Edges between nodes indicate flow of data
between stages, and each data element that flows through the edges is a \emph{primitive}.
Examples of common primitives are patches, vertices, triangles, and fragments.
Stages that
have no incoming edges are \emph{source} stages, and stages with no outgoing
edges are \emph{drain} stages.

Programmers divide each \piko{} pipeline stage into three \emph{phases}
(summarized in \tabref{phases} and \figref{dataflow}). The input to a
stage is a group or list of primitives, but
phases, much like OGL/D3D shaders, are programs that apply to a single
input element (e.g., a primitive or bin) or a small group thereof.
However, unlike shaders, phases belong in each stage of the pipeline,
and provide structural as well as functional information about a stage's
computation.
The first phase in a stage, \assignbin,
specifies how a primitive is mapped to a user-defined bin. The second
phase, \schedule, assigns bins to cores. The third phase, \process,
performs the actual functional computation for the stage on the primitives
in a bin. Allowing
the programmer to specify both how primitives are binned and how bins
are scheduled onto cores allows \pikoc{} to take advantage of spatial
locality.

Now, if we simply replace $n$ pipeline stages with $3n$ simpler phases and
invoke our \textsc{Baseline} implementation,
we would gain little benefit from this factorization. Fortunately, we
have identified and implemented several high-level optimizations on
stages and phases that make this factorization profitable.
We describe some of our optimizations in Section~\ref{sec:optimizations}.

As an example, \listref{kernels} shows the phases of a very simple fragment
shader stage. In the \assignbin{} stage, each fragment goes into a single
bin chosen based on its screen-space position. To maximally exploit the 
machine parallelism, \schedule{} requests the runtime to distribute bins in a load-balanced
fashion across the machine. \process{} then executes a simple pixel shader,
since the computation by now is well-distributed across all available cores.
For the full source code of this simple raster pipeline, please refer
to the supplementary materials. Now, let us describe each phase in
more detail.

\paragraph{\assignbin}

The first step for any incoming primitive is to identify the tile(s)
that it may influence or otherwise belongs to. Since this depends on
both the tile structure as well as the nature of computation in the
stage, the programmer is responsible for mapping primitives to bins.
Primitives are put in bins with the \texttt{assignToBin} function that
assigns a primitive to a bin. \listref{kernels} assigns an input fragment $f$
based on its screen-space position.

\begin{lstlisting}[float,caption={Example \piko{} routines for a
fragment shader pipeline stage and its corresponding pipeline RasterPipe.
In the listing, blue indicates \piko{}-specific keywords,
purple indicates user-defined objects, and sea-green indicates user-defined functions.
The template parameters to \textbf{Stage} are, in order: binSizeX, binSizeY,
threads per bin, incoming primitive type, and outgoing primitive
type. We specify a LoadBalance scheduler to take advantage of the many cores on the GPU.
} ,label=list:kernels,belowskip=-0.27in]{t}

class FragmentShaderStage :
    //  This stage has 8x8 pixel bins, and requests
    //  64 threads for each invocation. Input as
    //  well as output of this stage is a fragment.
    public Stage<8, 8, 64, piko_fragment, piko_fragment> {
  public:
    void assignBin(piko_fragment f) {
      int binID = getBinFromPosition(f.screenPos);
      this->assignToBin(f, binID);
    }

    void schedule(int binID) {
      specifySchedule(LOAD_BALANCE);
    }

    void process(piko_fragment f) {
      cvec3f material = gencvec3f(0.80f, 0.75f, 0.65f);
      cvec3f lightvec = normalize(gencvec3f(1,1,1));
      f.color = material * dot(f.normal, lightvec);
      this->emit(f,0);
    }
};
\end{lstlisting}

\paragraph{\schedule}
The best execution schedule for computation in a pipeline varies with 
stage, characteristics of the pipeline input, and target architectures.
Thus, it is natural to want to customize scheduling preferences in order
to retarget a pipeline to a different scenario. Furthermore, many pipelines 
impose constraints on the observable order in which
primitives are processed. In \piko{}, the programmer explicitly provides
such preference and constraints on how bins are scheduled on execution cores.
Specifically, once primitives are assigned into
bins, the \schedule{} phase allows the programmer to specify how and
when bins are scheduled onto cores. The input to \schedule{} is a
reference to a spatial bin, and the routine can chose to dispatch
computation for that bin, and if it does, it can also choose a specific
execution core or scheduling preference.

We also recognize two cases of special scheduling
constraints in the abstraction: the case where all bins from one stage
must complete processing before a subsequent stage can begin, and the
case where all primitives from one bin must complete processing before
any primitives in that bin can be processed by a subsequent stage.
\listref{kernels} shows an example of a \schedule{} phase that
schedules primitives to cores in a load-balanced
fashion.

Because of the variety of scheduling mechanisms and strategies on
different architectures, we expect \schedule{} phases to be
the most architecture-dependent of the three. For instance, a manycore GPU implementation
may wish to maximize utilization of cores by load balancing its computation, 
whereas a CPU might choose to schedule in chunks to preserve cache
peformance, and hybrid CPU-GPU may wish to
preferentially assign some tasks to a particular processor (CPU or GPU).

\schedule{} phases specify not only where computation will take place
but also when that computation will be launched. For instance, the
programmer may specify dependencies that must be satisfied before
launching computation for a bin. For example, an order-independent
compositor may only launch on a bin once all its fragments are
available, and a fragment shader stage may wait for a sufficiently large
batch of fragments to be ready before launching the shading computation.
Currently, our implementation \pikoc{} resolves such
constraints by adding barriers between stages, but a future implementation
might choose to dynamically resolve such dependencies.

\paragraph{\process}

While \assignbin{} defines how primitives are grouped for computation into
bins, and \schedule{} defines where and when that computation takes place,
the \process{} phase defines the typical functional role of the stage. The 
most natural
example for \process{} is a vertex or fragment shader, but \process{}
could be an intersection test, a depth resolver, a subdivision task, or any
other piece of logic that would typically form a standalone stage in a
conventional graphics pipeline description. The input to \process{} is
the primitive on which it should operate. Once a primitive is processed
and the output is ready, the output is sent to the next stage via the
\texttt{emit} keyword. \texttt{emit} takes the output and an ID that
specifies the next stage. In the graph analogy of nodes (pipeline stages),
the ID tells the current node which edge to traverse down toward the
next node.  Our notation is that \process{} \emph{emit}s from zero to many
primitives that are the input to the next stage or stages.

We expect that many \process{} phases will exhibit data parallelism
over the primitives. Thus, by default, the input to \process{} is a
single primitive. However, in some cases, a \process{} phase may be
better implemented using a different type of parallelism or may require
access to multiple primitives to work correctly. For these cases, we
provide a second version of \process{} that takes a list of primitives
as input. This option allows flexibility in how the phase utilizes
parallelism and caching, but it limits our ability to perform pipeline
optimizations like kernel fusion (discussed in \secref{kernel-fusion}).
It is also analogous to the categorization of graphics code into
\emph{pointwise} and \emph{groupwise} computations, as presented by 
Foley et al.~\shortcite{Foley:2011:SMC}.

\subsection{Programming Interface}

A developer of a \piko{} pipeline supplies a pipeline definition with each
stage separated into three phases: \assignbin, \schedule, and
\process{}. \pikoc{} analyzes the code to generate a pipeline skeleton
that contains information about the vital flow of the pipeline. From
the skeleton, \pikoc{} performs a synthesis stage where it merges
pipeline stages together to output an efficient set of kernels that
executes the original pipeline definition.  The optimizations performed
during synthesis, and different runtime implementations of the \piko{}
kernels, are described in detail in Section~\ref{sec:optimizations}
and Section~\ref{sec:implementation} respectively.

From the developer's perspective, one writes several pipeline stage
definitions; each stage has its own \assignbin, \schedule, and \process{}.
Then the developer writes a pipeline class that connects the pipeline stages
together. We express our stages in a simple C++-like language.

These input files are compiled by \pikoc{} into two files: a file
containing the target architecture kernel code, and a header file with a class
that connects the kernels to implement the pipeline. The developer creates 
an object of this class and calls the \texttt{run()} method to run the 
specified pipeline.

The most important architectural targets for \piko{} are multi-core 
CPU architectures and manycore GPUs\footnote{In this paper we
define a ``core'' as a hardware block with an independent program
counter rather than a SIMD lane; for instance, an NVIDIA streaming
multiprocessor (SM).}, and \pikoc{} is able to generate code for both.
In the future we also would like to extend its capabilities to
target clusters of CPUs and GPUs, and CPU-GPU hybrid architectures.

\subsection{Using Directives When Specifying Stages}
\label{sec:language}

\pikoc{} exposes several special keywords, which we call \emph{directives}, to 
help a developer directly express commonly-used yet complex implementation
preferences.
We have found that it is usually best for the developer to explicitly state
a particular preference, since it is often much easier to do so, and at the 
same time it helps enable optimizations which \pikoc{} might not have gathered
using static analysis. For instance, if the developer wishes to broadcast a 
primitive to all bins in the next stage, he can simply use \kw{AssignToAll}
in \assignbin. Directives act as compiler hints and further increase optimization 
potential. We summarize our directives in Table~\ref{tab:keywords} and discuss 
their use in Section~\ref{sec:optimizations}.

We combine these directives with the information that \pikoc{} derives
in its \emph{analysis} step to create what we call a \emph{pipeline
skeleton.} The skeleton is the input to \pikoc{}'s \emph{synthesis} step,
which we also describe in \secref{optimizations}.

\begin{table*}[t]
\begin{center}

\begin{tabular}{llp{1.12\columnwidth}}

\toprule
\textbf{Phase} & \textbf{Name}               & \textbf{Purpose}\\
\midrule
\multirow{4}{*}{
\assignbin{}}  & \kw{AssignPreviousBins}     & Assign incoming primitive to the same bin as in the previous stage\\
               & \kw{AssignToBoundingBox}    & Assign incoming primitive to bins based on its bounding box\\
               & \kw{AssignToAll}            & Assign incoming primitive to all bins\\
\midrule
\multirow{7}{*}{
\schedule{}}   & \kw{DirectMap}              & Statically schedule each bin to available cores in a round-robin fashion\\
               & \kw{LoadBalance}            & Dynamically schedule bins to available cores in a load-balanced fashion\\
               & \kw{Serialize}              & Schedule all bins to a single core for sequential execution\\
               & \kw{All}                    & Schedule a bin to all cores (used with \kw{tileSplitSize})\\
               & \kw{tileSplitSize}          & Size of chunks to split a bin across multiple cores (used with \kw{All})\\
               & \kw{EndStage(X)}            & Wait until stage \kw{X} is finished\\
               & \kw{EndBin}                 & Wait until the previous stage finishes processing the current bin\\
\bottomrule

\end{tabular}

\caption{The list of directives the programmer can specify to \piko{} during each phase. The directives provide basic structural information about the workflow and facilitate optimizations.}

\label{tab:keywords}
\end{center}
\end{table*}

\subsection{Expressing Common Pipeline Preferences}
\label{sec:common-pipelines}

We now present a few commonly encountered pipeline design strategies,
and how we interpret them in our abstraction:
\vspace{-1mm}

\paragraph{No Tiling} In cases where tiling is not a beneficial
choice, the simplest way to indicate it in \piko{} is to set bin sizes
of all stages to $0\times0$ (\pikoc{} translates it to the screen
size). Usually such pipelines (or stages) exhibit parallelism at the
per-primitive level. In \piko{}, we can use \kw{All} and 
\kw{tileSplitSize} in \schedule{} to specify the size of individual primitive-parallel
chunks.\vspace{-1mm}

\paragraph{Bucketing Renderer} Due to resource constraints, often the
best way to run a pipeline to completion is through a depth-first
processing of bins, that is, running the entire pipeline (or a sequence
of stages) over individual bins in serial order. In \piko{}, it is
easy to express this preference through the use of the \kw{All}
directive in \schedule{}, wherein each bin of a stage maps to all available cores.
Our synthesis scheme prioritizes depth-first processing in
such scenarios, preferring to complete as many stages for a bin before
processing the next bin. See \secref{basic-mapping} for details.
\vspace{-1mm}

\paragraph{Sort-Middle Tiled Renderer} A common design
methodology for forward renderers divides the pipeline into two
phases: world-space geometry processing and screen-space fragment
processing. Since \piko{} allows a different bin size for each stage,
we can simply use screen-sized bins with primitive-level parallelism
in the geometry phase, and smaller bins for the screen-space
processing.
\vspace{-1mm}

\paragraph{Use of Fixed-Function Hardware Blocks} Fixed-function
hardware accessible through CUDA or OpenCL (like texture fetch units)
is easily integrated into \piko{} using the mechanisms in those APIs.
However, in order to use standalone units like a hardware rasterizer
or tessellation unit that cannot be directly addressed, the best way
to abstract them in \piko{} is through a stage that implements a
single pass of an OGL/D3D pipeline. For example, a deferred rasterizer
could use OGL/D3D for the first stage, then a \piko{} stage to
implement the deferred shading pass.

\section{Pipeline Synthesis with \pikoc{}}
\label{sec:optimizations}

\pikoc{} is built on top of the LLVM compiler framework. Since \piko{}
pipelines are written using a subset of C++, \pikoc{} uses Clang,
the C and C++ frontend to LLVM, to compile pipeline source code into
LLVM IR\@. We further use Clang in \pikoc{}'s analysis step by walking the
abstract syntax tree (AST) that Clang generates from the source code.
From the AST, we are able to obtain the directives and infer the other
optimization information discussed previously, as well as determine how
pipeline stages are linked together. \pikoc{} adds this information to
the pipeline skeleton, which summarizes the pipeline and contains all the
information necessary for pipeline optimization.

\pikoc{} then performs pipeline synthesis in three steps. First,
we identify the order in which we want to launch individual stages
(\secref{pipeline-schedule}). Once we have this high-level stage
ordering, we optimize the organization of kernels to both maximize
producer-consumer locality and eliminate any redundant/unnecessary
computation (\secref{basic-mapping}). The result of this process is the
\emph{kernel mapping}: a scheduled sequence of kernels and the phases
that make up the computation inside each. Finally, we use the kernel
mapping to output two files that implement the pipeline: the kernel code
for the target architecture and a header file that contains host
code for setting up and executing the kernel code.

We follow typical convention for building complex applications on GPUs
using APIs like OpenCL and CUDA by instantiating a pipeline as a series
of kernels. Each kernel represents a machine-wide computation consisting
of parts of one or more pipeline stages. Rendering each frame consists
of launching a sequence of kernels scheduled by a host, or a CPU thread
in our case. Neighboring kernel instances do not share local memory,
e.g., caches or shared memory.

An alternative to multi-kernel design is to express the entire pipeline
as a single kernel, which manages pipeline computation via dynamic work
queues and uses a persistent-kernel approach~\cite{Aila:2009:UTE,Gupta:2012:ASO}
to efficiently schedule the computation. This is an attractive strategy
for implementation and has been used in OptiX, but we prefer the multi-kernel 
strategy for two reasons. First, efficient dynamic work-queues are complicated to 
implement on many core architectures and work best for a single, highly irregular stage.
 Second, the major advantages of 
dynamic work queues, including dynamic load balance and the ability to
capture producer-consumer locality, are already exposed to our implementation
through the optimizations we present in this section.

\label{sec:pikoc-status}

Currently, \pikoc{} targets two hardware architectures: multicore CPUs
and NVIDIA GPUs.  In addition to LLVM's many CPU backends, NVIDIA's
libNVVM compiles LLVM IR to PTX assembly code, which can then be
executed on NVIDIA GPUs using the CUDA driver
API\footnote{https://developer.nvidia.com/cuda-llvm-compiler}.
In the future, \pikoc{}'s LLVM integration will allow us to
easily integrate new back ends (e.g., LLVM backends for SPIR and HSAIL)
that will automatically target heterogeneous processors like Intel's Haswell
or AMD's Fusion. To integrate a new backend into \pikoc{}, we also need
to map all \piko{} API functions to their counterparts in the new backend
and create a new host code generator that can set up and launch
the pipeline on the new target.

\subsection{Scheduling Pipeline Execution} 
\label{sec:pipeline-schedule}

Given a set of stages arranged in a pipeline, in what order should we
run these stages? The \piko{} philosophy is to use the pipeline
skeleton with the programmer-specified directives to build a
schedule\footnote{Please note that the scheduling described in this
  section is distinct from the \schedule{} phase in the \piko{}
  abstraction. Scheduling here refers to the order in which we run kernels in a
  generated \piko{} pipeline.}
for these stages. Unlike GRAMPS~\cite{Sugerman:2009:GAP}, which takes
a dynamic approach to global scheduling of pipeline stages, we use a
largely static global schedule due to our multi-kernel design.

The most straightforward schedule is for a linear, feed-forward
pipeline, such as the OGL/D3D rasterization pipeline. In this case, we
schedule stages in descending order of their distance from the last
(drain) stage.

By default, a stage will run to completion before the next stage
begins. However, we deviate from this rule in two cases: when we fuse
kernels such that multiple stages are part of the same kernel (discussed
in \secref{kernel-fusion}), and when we launch stages for bins in a
depth-first fashion (e.g., chunking), where we prefer to complete
an entire bin before beginning another. We generate a depth-first
schedule when a stage specification directs the entire machine to
operate on a stage's bins in sequential order (e.g., by using the \kw{All}
directive). In this scenario, we continue to launch successive stages for
each bin as long as it is possible; we stop when we reach a stage that
either has a larger bin size than the current stage or has a dependency
that prohibits execution. In other words, when given the choice between
launching the same stage on another bin or launching the next stage on
the current bin, we choose the latter. This decision is similar to the
priorities expressed in Sugerman et al.~\shortcite{Sugerman:2009:GAP}. In
contrast to GRAMPS, our static schedule prefers launching stages farthest from
the drain first, but during any stripmining or depth-first tile traversal, we
prefer stages closer to the drain in the same fashion as the dynamic scheduler in
GRAMPS\@. This heuristic has the following advantage: when multiple branches are
feeding into the draining stage, finishing the
shorter branches before longer branches runs the risk of over-expanding the state.
Launching the stages farthest from the drain ensures that the stages have enough memory
to complete their computation.

More complex pipeline graph structures feature branches. With these,
we start by partitioning the pipeline into disjoint linear branches,
splitting at points of convergence, divergence, or explicit dependency
(e.g., \kw{EndStage}).
This method results in linear, distinct branches with no stage
overlap. Within each branch, we order stages using the simple
technique described above. However, in order to determine inter-branch
execution order, we sort all branches in descending order of
the distance-from-drain of the branch's starting stage. We attempt to
schedule branches in this order as long as all inter-branch
dependencies are contained within the already scheduled branches. If
we encounter a branch where this is not true, we skip it until its
dependencies are satisfied. Rasterization with a shadow map requires
this more complex branch ordering method; the branch of the pipeline
that generates the shadow map should be executed before the main
rasterization branch.

The final consideration when determining stage execution order is
managing pipelines with cycles. For non-cyclic pipelines, we can
statically determine stage execution ordering, but cycles create a
dynamic aspect because we often do not know at compile time how many
times the cycle will execute. For cycles that occur within a single stage
(e.g., Reyes's \stage{Split} in Section~\ref{sec:evaluation}), we repeatedly launch the same stage until
the cycle completes. We acknowledge that launching a single kernel with a dynamic work queue is a 
better solution in this case, but \pikoc{} doesn't currently support that.
 Multi-stage cycles (e.g., the trace loop in a ray
tracer) pose a bigger stage ordering challenge. In the case where a stage
receives input from multiple stages, at least one of which is not part
of a cycle containing the current stage, we allow the current stage to
execute (as long as any other dependencies have been met). Furthermore,
by identifying the stages that loop back to previously executed stages,
we can explicitly determine which portions of the pipeline should be repeated.

Please refer to the supplementary material for some example pipelines
and their stage execution order.

\subsection{Pipeline Optimizations}
\label{sec:basic-mapping}

The next step in generating the kernel mapping for a pipeline is
determining the contents of each kernel. We begin with a basic,
conservative division of stage phases into kernels such that each
kernel contains three phases: the current stage's \schedule{} and
\process{} phases, and the next stage's \assignbin{} phase. This
structure realizes the simple, inefficient \textsc{Baseline} in which each kernel
fetches its bins, schedules them onto cores per \schedule{}, executes
\process{} on them, and writes the output to the next stage's bins using
the latter's \assignbin{}. The purpose of \pikoc's optimization step
is to use static analysis and programmer-specified directives to find
architecture-independent optimization opportunities. We discuss these
optimizations below.

\subsubsection{Kernel Fusion} 
\label{sec:kernel-fusion}

Combining two kernels into one---``kernel fusion''---both reduces
kernel overhead and allows an implementation to exploit
producer-consumer locality between the kernels.

\begin{center}
\includegraphics[width=2.3in, clip, trim=7mm 0mm 0mm 0mm]
{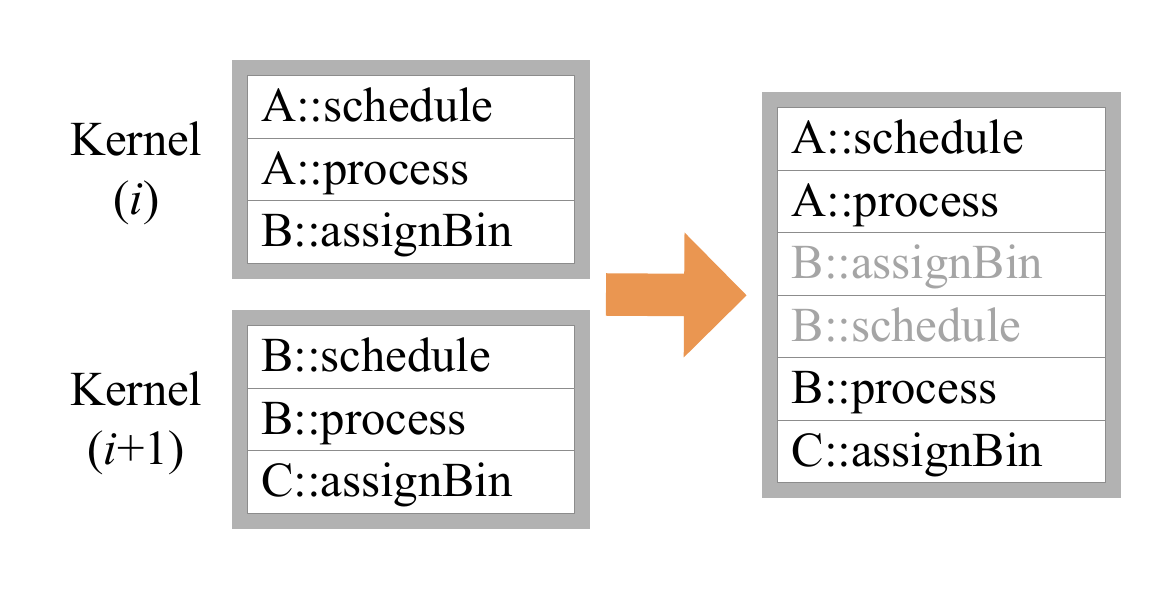}
\end{center}

\paragraph{Opportunity}

The simplest case for kernel fusion is when two subsequent stages (a)
have the same bin size, (b) map primitives to the same bins, (c) have
no dependencies between them, (d) each receive input from only one
stage and output to only one stage, and (e) both have \schedule{}
phases that map execution to the same core. For example, a rasterization pipeline's
\stage{Fragment Shading} and \stage{Depth Test} stages can be fused.
If requirements are met, a primitive can proceed from one stage
to the next immediately and trivially, so we fuse these two stages
into one kernel. These constraints can be relaxed in certain cases
(such as a \stage{EndBin} dependency, discussed below),
allowing for more kernel fusion opportunities. We anticipate more
complicated cases where kernel fusion is possible but difficult to
detect; however, even detecting only the simple case above is highly
profitable.

\paragraph{Implementation}

Two stages, A and B, can be fused by having A's emit statements call
B's process phase directly. We can also fuse more than two stages
using the same approach.

\subsubsection{\schedule{} Optimization} 
\label{sec:schedule-optimization}

While we allow a user to express arbitrary logic in a \schedule{}
routine, we observe that most common patterns of scheduler design can
be reduced to simpler and more efficient versions. Two prominent cases
include:

\vspace{3mm}
\textbf{Pre-Scheduling}

\begin{center}
\includegraphics[width=2.3in, clip, trim= 7mm 0mm 0mm 0mm]
{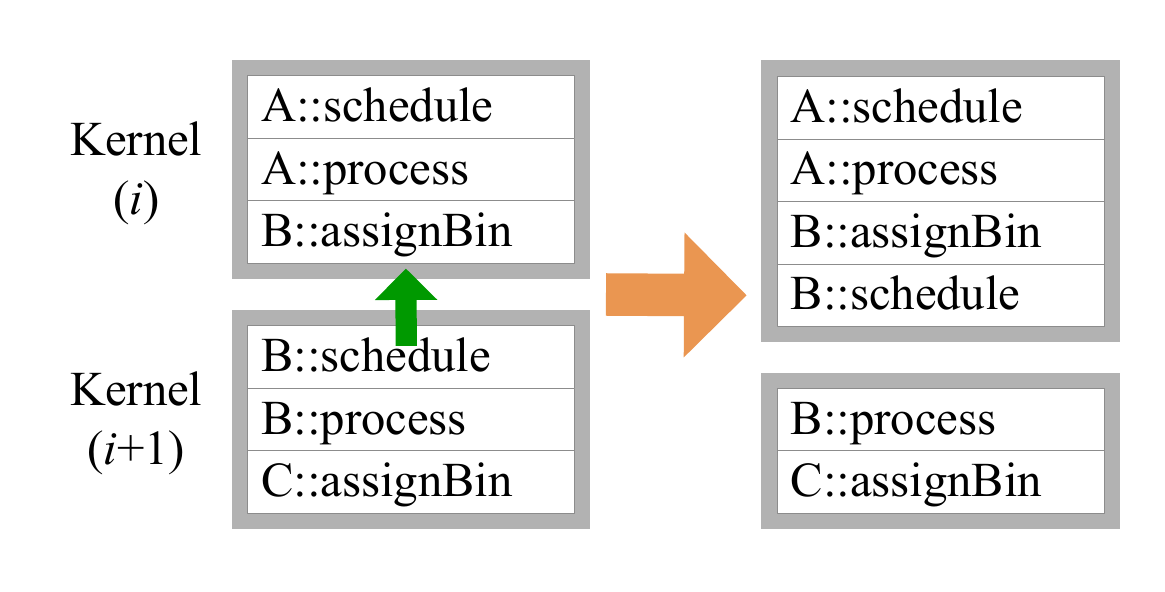}
\end{center}

\paragraph{Opportunity}
For many \schedule{} phases, core selection is either static or
deterministic given the incoming bin ID (specifically, when
\kw{DirectMap}, \kw{Serialize}, or \kw{All}
are used). In these scenarios, we can pre-calculate the target core ID
even before \schedule{} is ready for execution (i.e., before all
dependencies have been met). This both eliminates some runtime work
and provides the opportunity to run certain tasks (such as data
allocation on heterogeneous implementations) before a stage is ready
to execute.

\paragraph{Implementation}
The optimizer detects the pre-scheduling optimization by
identifying one of the three aforementioned \schedule{}
directives. This optimization allows us to move a given stage's
\schedule{} phase into the same kernel as its \assignbin{}
phase so that core selection happens sooner and so that other
implementation-specific benefits can be exploited.

\vspace{3mm}
\textbf{Schedule Elimination}

\begin{center}
\includegraphics[width=2.3in, clip, trim= 7mm 0mm 0mm 0mm]
{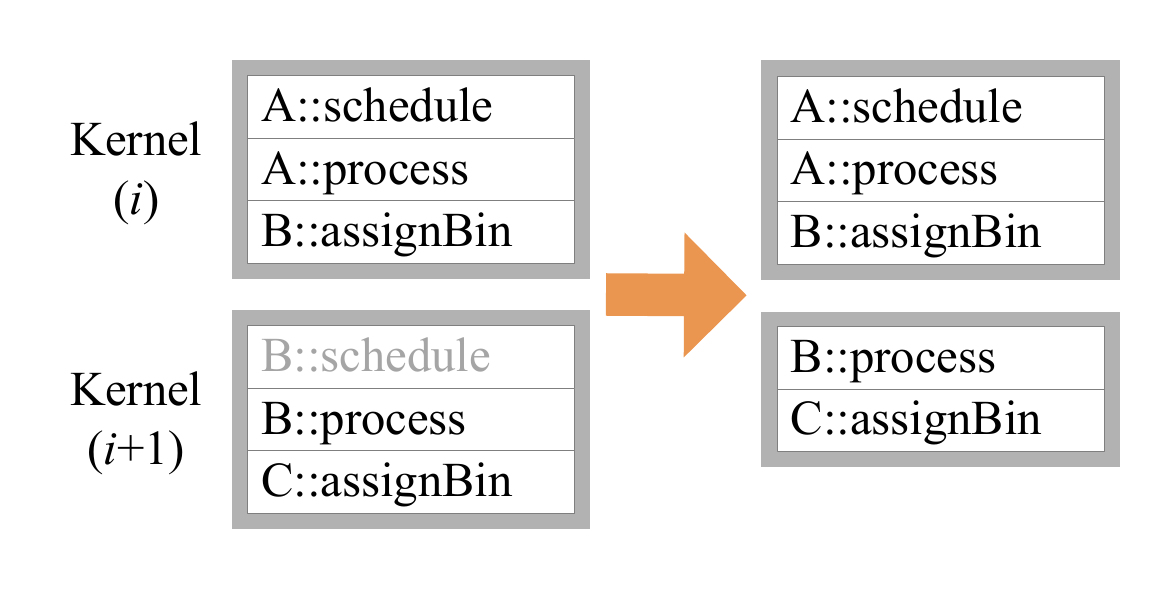}
\end{center}

\paragraph{Opportunity}
Modern parallel architectures often support a highly efficient
hardware scheduler that offers a reasonably fair allocation of
work to computational cores. Despite the limited customizability
of such a scheduler, we utilize its capabilities whenever
it matches a pipeline's requirements. For instance, if a
designer requests bins of a fragment shader to be scheduled in
a load-balanced fashion (e.g., using the \kw{LoadBalance}
directive), we can simply offload this task to the hardware
scheduler by presenting each bin as an independent unit of work
(e.g., a CUDA block or OpenCL workgroup).

\paragraph{Implementation}
When the optimizer identifies a stage using the
\kw{LoadBalance} directive, it removes that stage's
\schedule{} phase in favor of letting the hardware scheduler
allocate the workload.

\subsubsection{Static Dependency Resolution} 

\begin{center}

\includegraphics[width=2.5in, clip, trim= 7mm 12mm 0mm 0mm]
{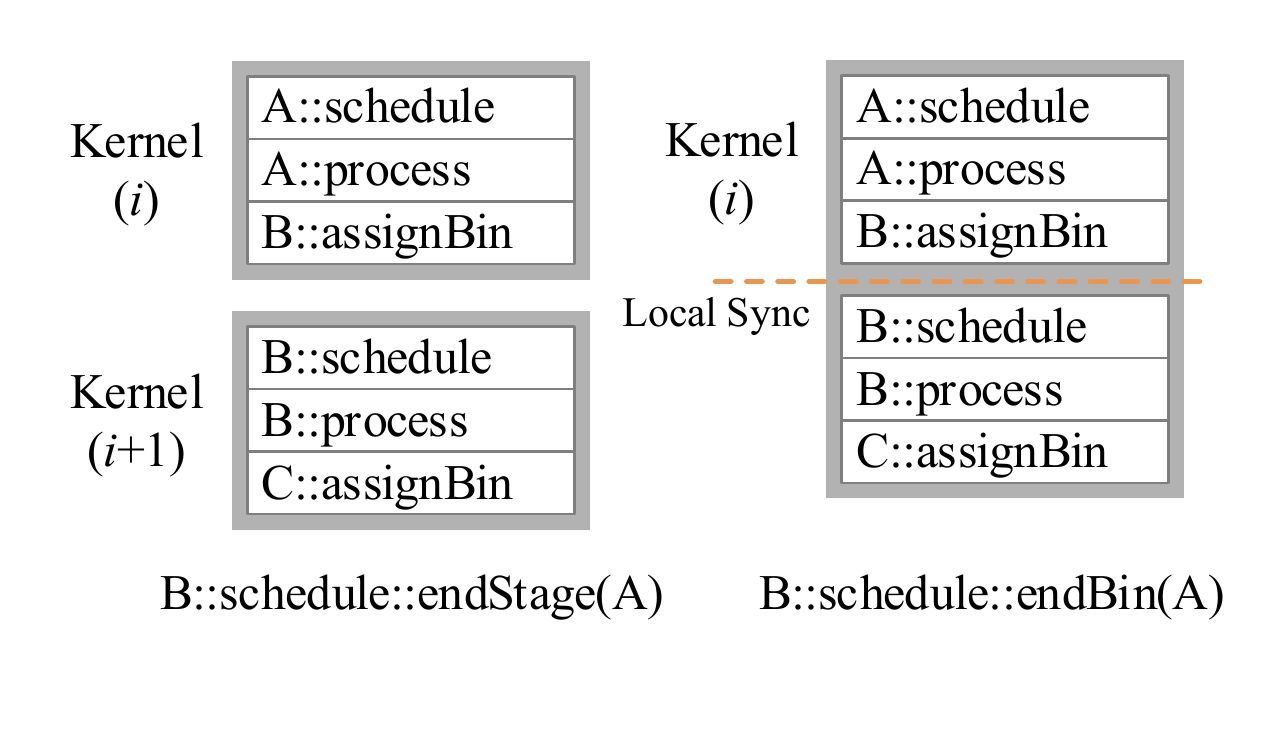}
\end{center}

\paragraph{Opportunity}

The previous optimizations allowed us to statically resolve core
assignment. Here we also optimize for static resolution of
dependencies. The simplest form of dependencies are those that request
completion of an upstream stage (e.g., the \kw{EndStage}
directive) or the completion of a bin from the previous stage (e.g., the
\kw{EndBin} directive). The former dependency occurs in
rasterization pipelines with shadow mapping, where the \stage{Fragment
  Shade} stage cannot proceed until the pipeline has finished
generating the shadow map (specifically, the shadow map's
\stage{Composite} stage). The latter dependency occurs when
synchronization is required between two stages, but the requirement is
spatially localized (e.g., between the \stage{Depth Test} and
\stage{Composite} stages in a rasterization pipeline with 
order-independent transparency).

\paragraph{Implementation}
We interpret \kw{EndStage} as a global synchronization
construct and, thus, prohibit any kernel fusion with a previous stage.
By placing a kernel break between stages, we enforce the
\kw{EndStage} dependency because once a kernel has finished
running, the stage(s) associated with that kernel are complete.

In contrast, \kw{EndBin} denotes a local synchronization,
so we allow kernel fusion and place a local synchronization
within the kernel between stages. However, this strategy only
works if a bin is not split across multiple cores. If a bin is
split, we fall back to global synchronization.

\subsubsection{Single-Stage \process{} Optimizations}

Currently, we treat \process{} stages as architecture-independent. In
general, this is a reasonable assumption for graphics pipelines.
However, we have noted some specific scenarios where
architecture-dependent \process{} routines might be desirable. For
instance, with sufficient local storage and small enough bins, a
particular architecture might be able to instantiate an on-chip depth
buffer, or with a fast global read-only storage, lookup-table-based
rasterizers become possible. Exploring architecture-dependent
\process{} stages is an interesting area of future work.

\section{Runtime Implementation}
\label{sec:implementation}
\label{sec:common-implementation}

We designed \piko{} to target multiple architectures,
and we currently focus on two distinct targets: a multicore CPU
and a manycore GPU\@.
Certain aspects of our runtime design span both architectures. The
uniformity in these decisions provides a good context for comparing
differences between the two architectures. The degree of impact of 
optimizations in \secref{basic-mapping} generally
varies between architectures, and that helps us tweak pipeline
specifications to exploit architectural strengths. Along with using
multi-kernel implementations, our runtimes also share the following
characteristics:

\paragraph{Bin Management} For both architectures, we consider a
simple data structure for storing bins: each stage maintains a list of
bins, each of which is a list of primitives belonging to the
corresponding bin. Currently, both runtimes
use atomic operations to read and write to bins. However, using
prefix sums for updating bins while maintaining primitive order is a
potentially interesting alternative.

\paragraph{Work-Group Organization} In order to accommodate the
most common scheduling directives of static and dynamic load balance, we
simply package execution work groups into CPU threads/CUDA blocks such that they
respect the directives we described in Section~\ref{sec:language}:

\begin{description}

\item[LoadBalance] As discussed in \secref{schedule-optimization}, for
  dynamic load balancing \pikoc{} simply relies on the hardware
  scheduler for fair allocation of work. Each bin is assigned to
  exactly one CPU thread/CUDA block, which is then scheduled for
  execution by the hardware scheduler.

\item[DirectMap] While we cannot guarantee that a specific computation
  will run on a specific hardware core, here \pikoc{} packages
  multiple pieces of computation---for example, multiple bins---together 
  as a single unit to ensure that they will all run
  on the same physical core.

\end{description}

\label{sec:backends}

\Piko{} is designed to target multiple architectures by primarily changing
the implementation of the \schedule{} phase of stages. Due to the intrinsic
architectural differences between different hardware targets, the \Piko{}
runtime implementation for each target must exploit the unique architectural
characteristics of that target in order to obtain efficient pipeline implementations.
Below are some architecture-specific runtime implementation details.

\paragraph{Multicore CPU}
In the most common case, a bin will be assigned to
a single CPU thread. When this mapping occurs, we can manage the bins
without using atomic operations. Each bin will then be processed serially
by the CPU thread. 

Generally, we tend to prefer \kw{DirectMap} \schedule{}s. This
scheduling directive often preserves producer-consumer locality by
mapping corresponding bins in different stages to the same hardware
core. Today's powerful CPU cache hierarchies allow us to better
exploit this locality.

\paragraph{NVIDIA GPU}
High-end discrete GPUs have a large number of
wide-SIMD cores. We thus prioritize supplying large amounts of work to
the GPU and ensuring that work is relatively uniform. In specifying
our pipelines, we generally prefer \schedule{}s that use the efficient,
hardware-assisted \kw{LoadBalance} directive whenever appropriate.

Because we expose a threads-per-bin choice to the user when defining a
stage, the user can exploit knowledge of the pipeline and/or expected
primitive distribution to maximize efficiency. For example, if the
user expects many bins to have few primitives in them, then the user
can specify a small value for threads-per-bin so that multiple bins get
mapped to the same GPU core. This way, we are able to exploit locality
within a single bin, but at the same time we avoid losing performance
when bins do not have a large number of primitives.

\section{Evaluation}
\label{sec:evaluation}

\subsection{\Piko{} Pipeline Implementations}
\label{sec:evaluation-implementations}

In this section, we evaluate performance for two specific pipeline
implementations, described below---rasterization and Reyes---but the
\piko{} abstraction can effectively express a range of other pipelines
as well. In the supplementary material, we describe several additional
\piko{} pipelines, including a triangle rasterization pipeline with
deferred shading, a particle renderer, and and a ray tracer.

\paragraph{\textsc{Baseline} Rasterizer}
To understand how one can use \piko{} to design an efficient graphics pipeline, we begin by presenting
a \piko{} implementation of our \textsc{Baseline} triangle rasterizer. This pipeline consist of 5 stages
connected linearly: \stage{Vertex Shader}, \stage{Rasterizer}, \stage{Fragment Shader},
\stage{Depth Test}, and \stage{Composite}.  Each of these stages will use full-screen bins, which
means that they will not make use of spatial binning.  The \schedule{} phase for each stage
will request a \kw{LoadBalance} scheduler, which will result in each stage being mapped to its own
kernel. Thus, we are left with a rasterizer that runs each stage, one-at-a-time, to completion and
makes use of neither spatial nor producer-consumer locality. When we run this naive pipeline
implementation, the performance leaves much to be desired. We will see how we can improve
performance using \piko{} optimizations in Section~\ref{sec:evaluation-designAlternatives}.

\paragraph{Reyes}
As another example pipeline, let us explore a \piko{} implementation of a Reyes micropolygon renderer.
For our implementation, we split
the rendering into four pipeline stages: \stage{Split}, \stage{Dice},
\stage{Sample}, and \stage{Shade}. One of the biggest differences between Reyes and
a forward raster pipeline is that the \stage{Split} stage in Reyes is irregular
in both execution and data. Bezier patches may go through an unbounded number of
splits; each split may emit primitives that must be split again (\stage{Split}),
or instead diced (\stage{Dice}). These irregularities combined make Reyes
difficult to implement efficiently on a GPU\@. Previous GPU implementations of
Reyes required significant amounts of low-level, processor-specific code, such
as a custom software
scheduler~\cite{Patney:2008:RRA,Zhou:2009:RIR,Tzeng:2010:TMF,Weber:2015:PRA}.

In contrast, we represent \stage{Split} in \piko{} with only a few lines of
code. \stage{Split} is a self-loop stage with two output channels: one back to
itself, and the other to \stage{Dice}. \stage{Split}'s \schedule{} stage
performs the split operation and depending on the need for more splitting,
writes its output to one of the two output channels. \stage{Dice} takes in
Bezier patches as input and outputs diced micropolygons from the input patch.
Both \stage{Dice} and \stage{Sample} closely follow the GPU algorithm described
by Patney et al.~\shortcite{Patney:2008:RRA} but without its implementation
complexity. \stage{Shade} uses a diffuse shading model to color in the final
pixels.

\stage{Split} and \stage{Dice} follow a \kw{LoadBalance} scheme for scheduling
work with fullscreen bins. These bins do not map directly to screen space as
Bezier patches are not tested for screen coverage. Instead, in these stages, the
purpose of the bins is to help distribute work evenly. Since \stage{Sample} does
test for screen coverage, its bins partition the screen evenly into $32 \times
32$ bins. \stage{Shade} uses a \kw{DirectMap} scheme to ensure that generated
fragments from \stage{Sample} can be consumed quickly. To avoid the explosion of
memory typical of Reyes implementations, our
implementation strip-mines the initial set of patches via a tweakable knob
so that any single pass will fit within the GPU's available resources.

\begin{center}\ding{167}\end{center}

Since a \piko{} pipeline is written as a series of separate stages, we
can reuse these stages in other pipelines. For instance, the
\stage{Shade} stage in the Reyes pipeline is nearly identical to the
\stage{Fragment Shader} stage in the raster pipeline.  Furthermore,
since \piko{} factors out programmable binning into the \kw{AssignBin}
phase, we can also share binning logic between stages.  In Reyes, both
\stage{Split} and \stage{Dice} use the same round-robin \kw{AssignBin}
routine to ensure an even distribution of B{\'e}zier patches. The
\stage{Vertex Shader} stage of our binned rasterization pipeline
(described in \secref{strawmanWithBinning}) uses this \kw{AssignBin}
routine as well.  In addition, Reyes's \stage{Sample} stage uses the
same \kw{AssignBin} as the rasterizer's \stage{Rasterizer} stage,
since these two stages perform similar operations of screen-primitive
intersection.  Being able to reuse code across multiple pipelines and
stages is a key strength of \piko{}. Users can easily prototype and
develop new pipelines by connecting existing pipeline stages together.

\subsection{\piko{} Lets Us Easily Explore Design Alternatives}
\label{sec:evaluation-designAlternatives}

\begin{figure}[t]
\begin{center}
\begin{tabular}{rl}
\includegraphics[width=1.4in]{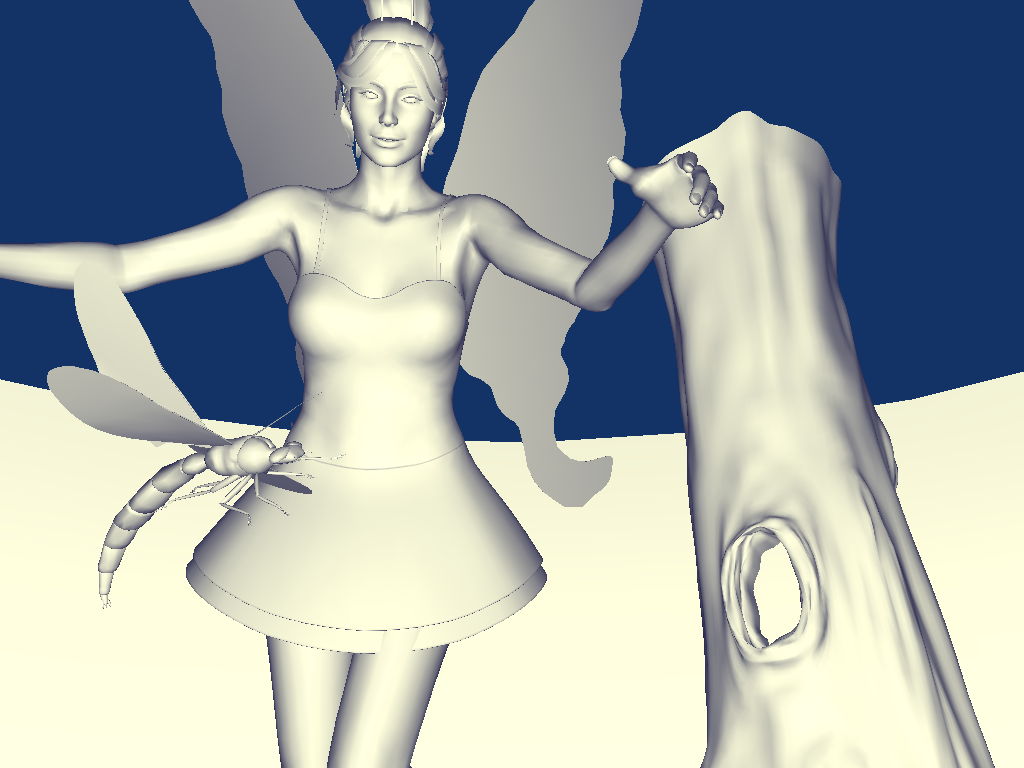} &
\includegraphics[width=1.4in]{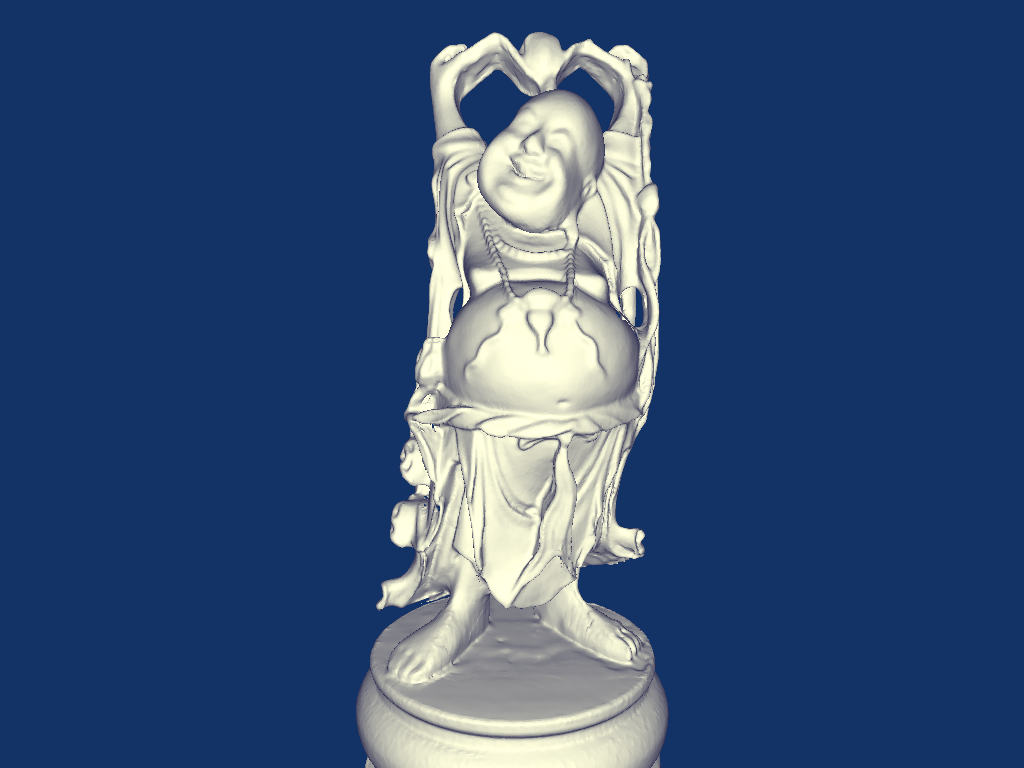}\\
\includegraphics[width=1.4in] {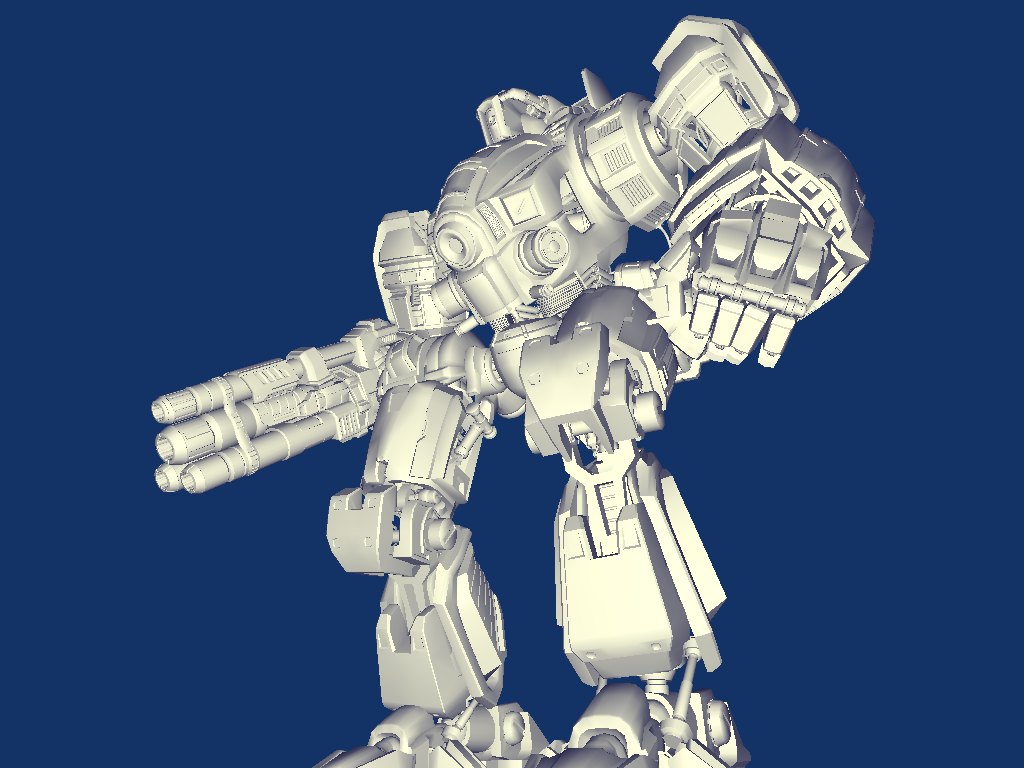} &
\includegraphics[width=1.4in] {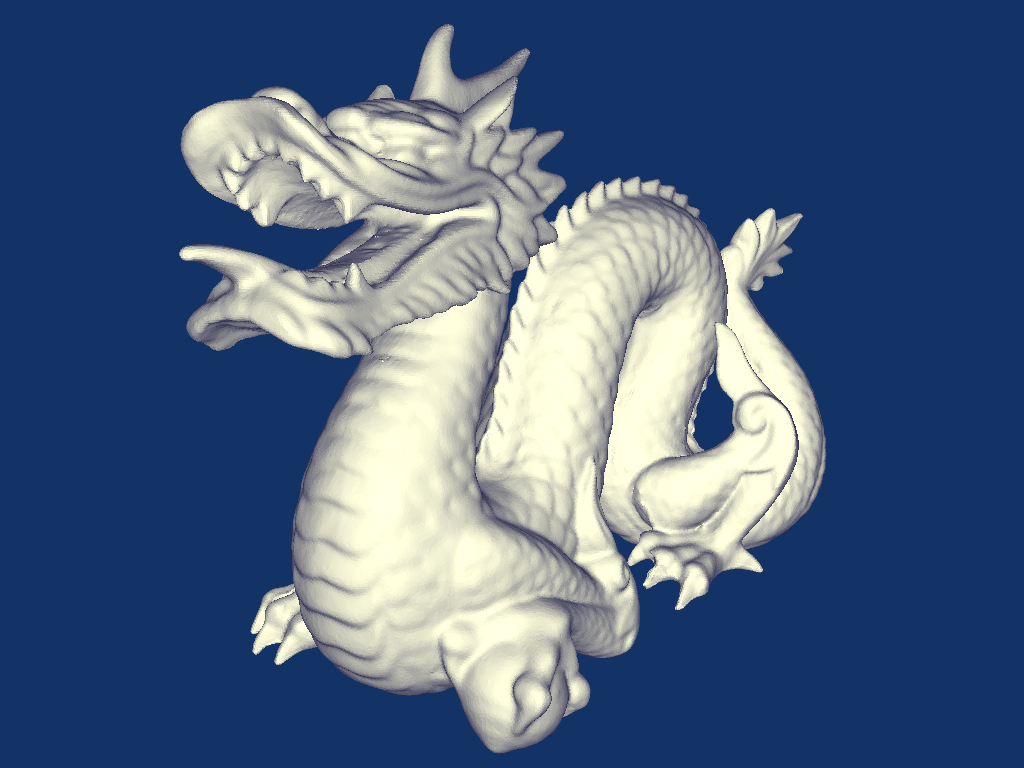}
\end{tabular}

\caption{We use the above scenes for evaluating characteristics of our
rasterizer implementations. Fairy Forest (top-left) is a scene with 174K
triangles with many small and large triangles. Buddha (top-right) is a
scene with 1.1M very small triangles. Mecha (bottom-left) has 254K small- to 
medium-sized triangles, and Dragon (bottom-right) contains 871K small triangles.
All tests were performed at a resolution of 1024$\times$768.}

\label{fig:rastpics}

\end{center}
\end{figure}

The performance of a graphics pipeline can depend heavily on the scene
being rendered. Scenes can differ in triangle size, count, and distribution,
as well as the complexity of the shaders used to render the scene.
Furthermore, different target architectures vary greatly in their design,
often requiring modification of the pipeline design in order to achieve
optimal performance on different architectures.
Now we will walk through a design
exercise, using the Fairy Forest and Buddha scenes (\figref{rastpics}), and show
how simple changes to \piko{} pipelines can allow both exploration and
optimization of design alternatives.

\begin{figure*}[t]
\begin{center}
\begin{subfigure}[b]{0.31\textwidth}
\includegraphics[width=\textwidth, clip, trim=0.8in 1.0in 1.1in 0.9in]
  {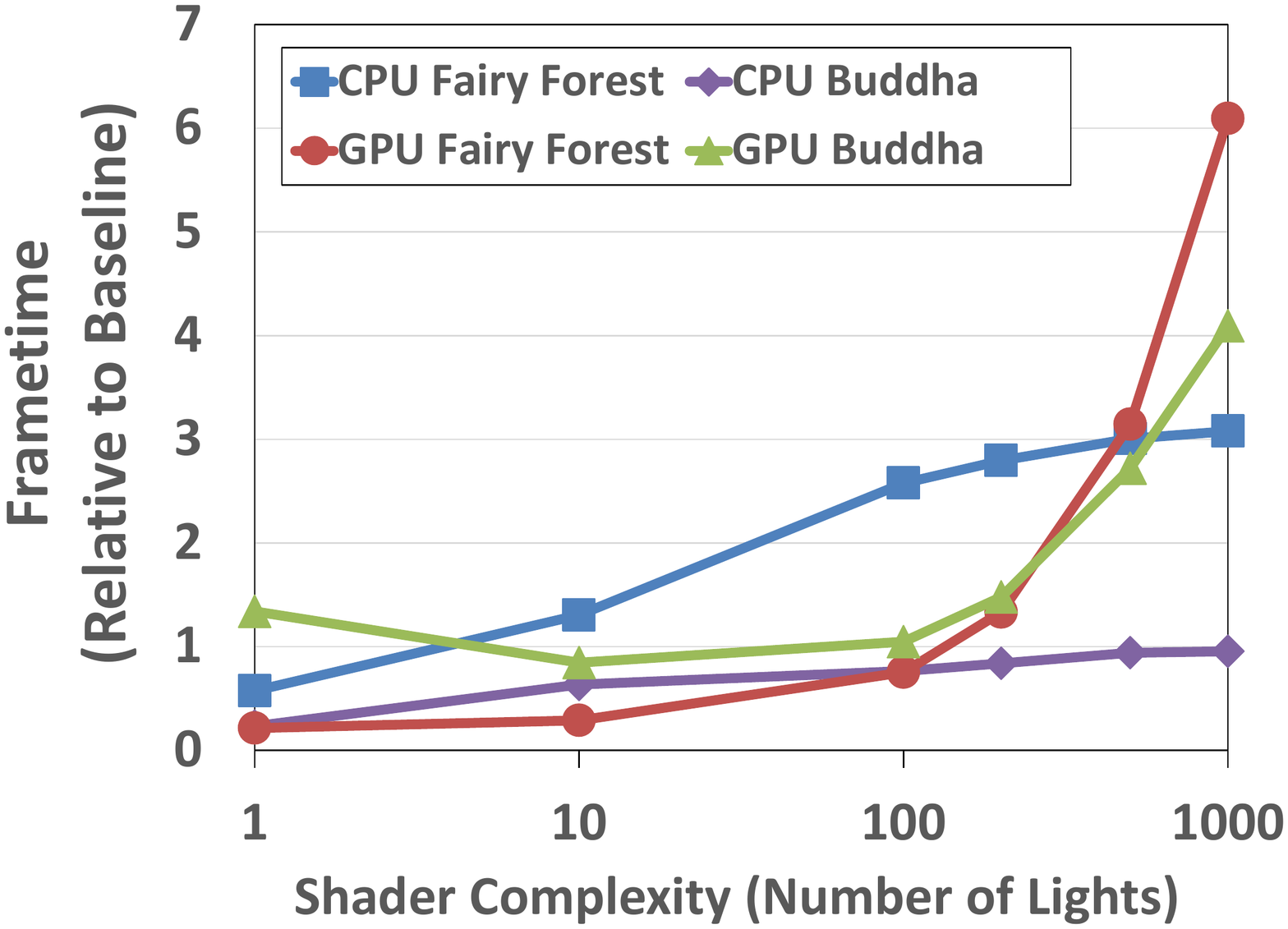}
\caption{FreePipe-style Aggressive Stage Fusion}
\label{fig:freepipe-strawman}
\end{subfigure}
\hspace*{\fill}
\begin{subfigure}[b]{0.31\textwidth}
\includegraphics[width=\textwidth, clip, trim=0.8in 1.0in 1.1in 0.9in]
  {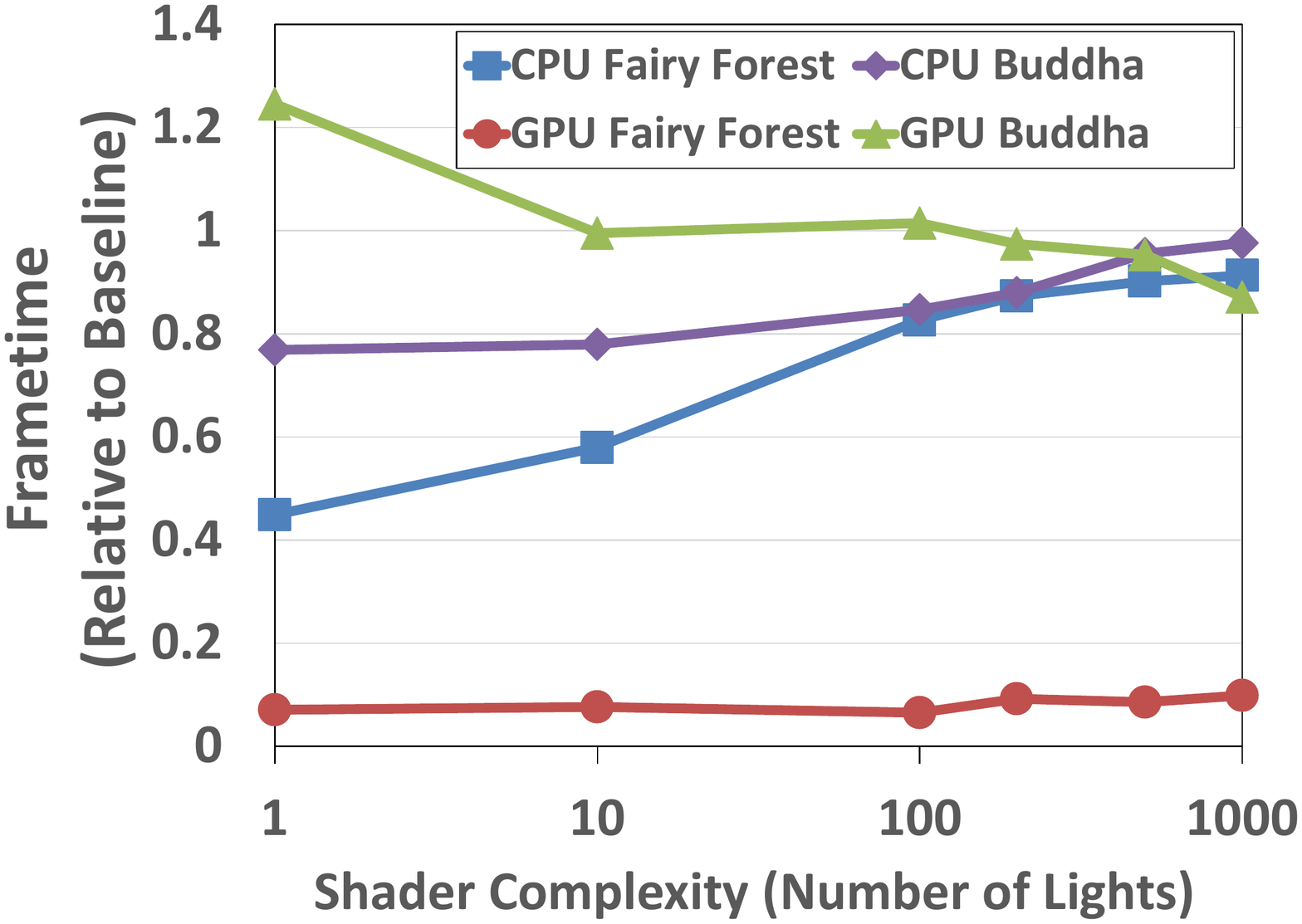}
\caption{Spatial Binning}
\label{fig:loadbalance-strawman}
\end{subfigure}
\hspace*{\fill}
\begin{subfigure}[b]{0.31\textwidth}
\includegraphics[width=\textwidth, clip, trim=0.8in 1.0in 1.1in 0.9in]
  {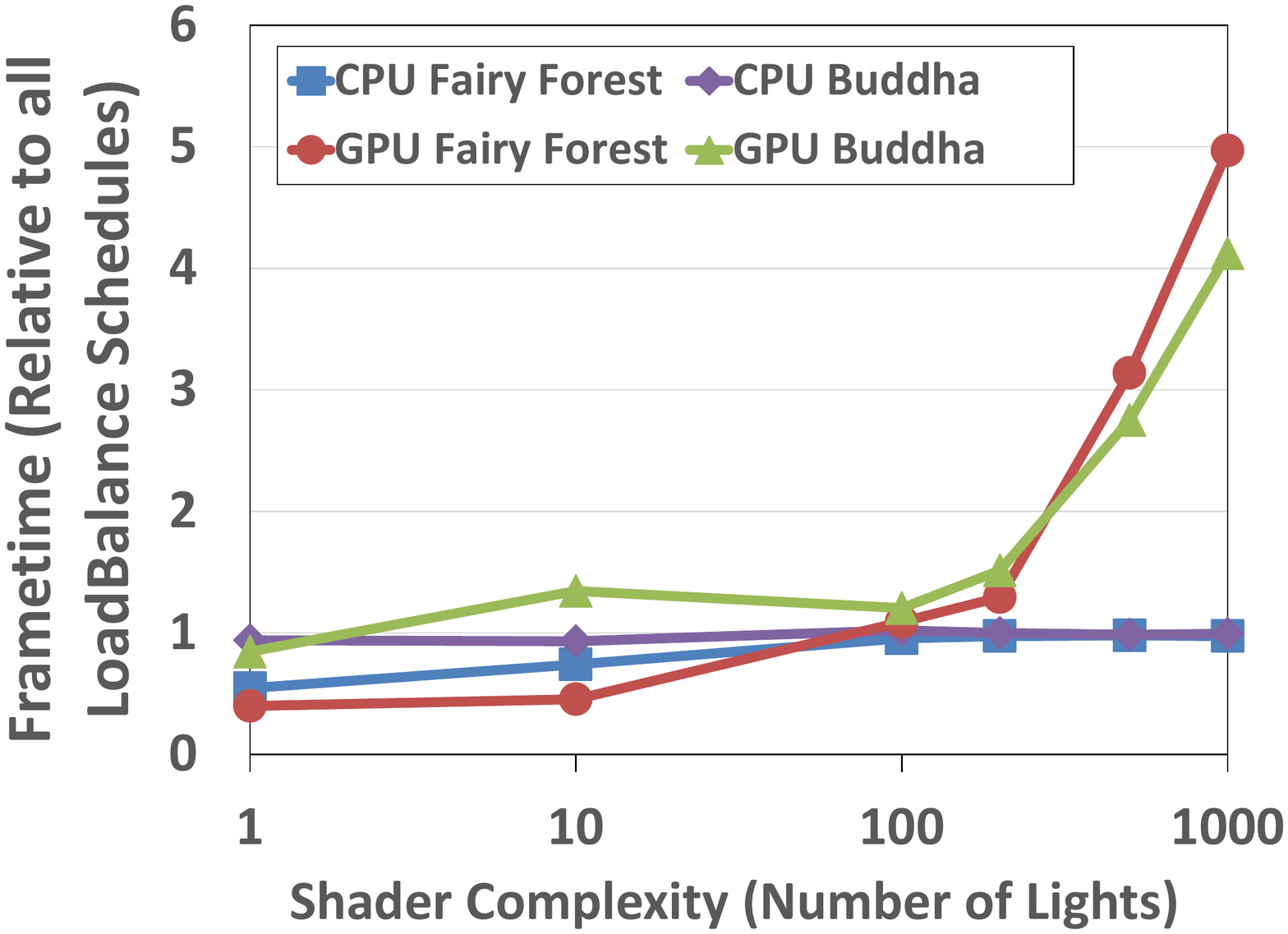}
\caption{Binning with Stage Fusion}
\label{fig:locality-loadbalance}
\end{subfigure}
\caption{Impact of various \piko{} configurations on the rendering performance of a rasterization pipeline. Frametimes are relative to \textsc{Baseline} for (a) and (b) and relative to using all \kw{LoadBalance} schedules for (c); lower is better. Shader complexity was obtained by varying the number of lights in the illumination computation. (a)~Relative to no fusion, FreePipe-style aggressive stage fusion is increasingly inefficient as shading complexity increases. However, it is a good choice for simple shaders. (b)~In almost all cases, using spatial binning results in a performance improvement. Because the triangles in the Buddha scene are small and regularly distributed, the benefits of binning are overshadowed by the overhead of assigning these many small triangles to bins. (c)~Compared to using all \kw{LoadBalance} schedules, the impact of fusing the \stage{Rasterization} and \stage{Fragment Shader} stages depends heavily on the target architecture. Both versions benefit from spatial binning, but only the fused case leverages producer-consumer locality. On the CPU, the cache hierarchy allows for better exploitation of this locality. However, on the GPU, the resultant loss in load-balancing due to the fusion greatly harms performance. \piko{} lets us quickly explore these design alternatives, and many others, without rewriting any pipeline stages.}
\end{center}
\end{figure*}

\subsubsection{Changing \textsc{Baseline}'s Scheduler}

Our \textsc{Baseline} pipeline separates each stage into separate kernels; it
leverages neither spatial nor producer-consumer locality, but
dynamically load-balances all primitives in each stage by using the
\kw{LoadBalance} scheduling directive. Let's instead consider a
pipeline design that assumes primitives are statically load-balanced
and optimizes for producer-consumer locality.  If we now specify the
\kw{DirectMap} scheduler, \pikoc{} aggressively fuses all stages
together into a single kernel; this simple change results in an
implementation faithful to the FreePipe design~\cite{Liu:2010:FAP}.

\figref{freepipe-strawman} shows how the relative performance of this
single-kernel pipeline varies with varying pixel shader complexity.
As shader complexity increases, the computation time of
shading a primitive significantly outweighs the time spent
loading the primitive from and storing it to memory.
Thus, the effects of poor load-balancing in the FreePipe design
become apparent because many of the cores of the hardware target will idle
while waiting for a few cores to finish their workloads.
For simple shaders, the memory bandwidth requirement overshadows
the actual computation time, so FreePipe's ability to preserve
producer-consumer locality becomes a primary concern.
This difference is particularly evident when running the pipeline on the GPU,
where load balancing is crucial to keep the
device's computation cores saturated with work. \piko{} lets us
quickly explore this tradeoff by only changing
the \schedule{}s of the pipeline stages.

\subsubsection{Adding Binning to \textsc{Baseline}}
\label{sec:strawmanWithBinning}

Neither \textsc{Baseline} nor the FreePipe designs exploit the spatial
locality that we argue is critical to programmable graphics pipelines.
Thus, let's return to the \kw{LoadBalance} \schedule{}s and apply that
schedule to a binned pipeline; \piko{} allows us to change the bin
sizes of each stage by simply changing the template parameters, as
shown in \listref{kernels}. By rapidly experimenting with different
bin sizes, we were able to obtain a speedup in almost all cases
(\figref{loadbalance-strawman}). We achieve a significant speedup on
the GPU using the Fairy Forest scene. However, the overhead of binning
results in a performance loss on the Buddha scene because the
triangles are small and similarly sized. Thus, a naive, non-binned
distribution of work suffices for this scene, and binning provides no
benefits. On the CPU, a more capable cache hierarchy means that adding
bins to our scenes is less useful, but we still achieve some speedup
nonetheless. For \piko{} users, small changes in pipeline descriptions
allows them to quickly make significant changes in pipeline
configuration or differentiate between different architectures.

\subsubsection{Exploring Binning with Scheduler Variations}

\piko{} also lets us easily combine \schedule{} optimizations with
binning. For example, using a \kw{DirectMap} schedule for the
\stage{Rasterizer} and \stage{Fragment Shader} stages means that these
two stages can be automatically fused together by \pikoc{}.
For GPUs in particular, this is a profitable optimization for low
shader complexity (\figref{locality-loadbalance}).

Through this exploration, we have shown that design decisions that
prove advantageous on one scenario or hardware target can actually
harm performance on a different one. To generalize, the complex cache
hierarchies on modern CPUs allow for better exploitation of spatial
and producer-consumer locality, whereas the wide SIMD processors and
hardware work distributors on modern GPUs can better utilize
load-balanced implementations. Using \piko{}, we can quickly make
these changes to optimize a pipeline for multiple targets. \pikoc{}
does all of the heavy lifting in restructuring the pipeline and
generating executable code, allowing \piko{} users to spend their time
experimenting with different design decisions, rather than having to
implement each design manually.

\subsection{\piko{} Delivers Real-Time Performance}
\label{sec:evaluation-performance}

As we have demonstrated, \piko{} allows pipeline writers to explore design trade-offs quickly
in order to discover efficient implementations. In this section, we compare our \piko{} triangle
rasterization pipeline and Reyes micropolygon pipeline against their respective state-of-the-art
implementations to show that \piko{} pipelines can, in fact, achieve respectable performance.
We ran our \piko{} pipelines on an NVIDIA Tesla K40c GPU\@.

\subsubsection{Triangle Rasterization}

We built a high-performance GPU triangle rasterizer using \piko{}. Our
implementation inherits many ideas from
cudaraster~\cite{Laine:2011:HSR}, but our programming model helps us separately express
the algorithmic and optimization concerns. \tabref{cudaraster} compares the performance
of \piko{}'s rasterizer
against cudaraster.  Across several test
scenes (\figref{rastpics}), we find that cudaraster is approximately 3--6$\times$ faster than \piko{} rasterizer.

\begin{table}
  \centering

    \begin{tabular}{lccc}
      \toprule
      Scene                 & cudaraster & \piko{} raster & Relative\\
                            & (ms/frame) & (ms/frame)     & Performance\\
      \midrule
      Fairy Forest          & 1.58       & 8.80           &  5.57$\times$\\
      Buddha                & 2.63       & 11.20           &  4.26$\times$\\
      Mecha                 & 1.91       & 6.40           &  3.35$\times$\\
      Dragon                & 2.58       & 10.20          &  3.95$\times$\\
      \bottomrule
    \end{tabular}

    \caption{Performance comparison of an optimized \piko{} rasterizer against cudaraster, the current state-of-the-art in software rasterization on GPUs. We find that a \piko{}-generated rasterizer is about 3--6$\times$ slower than cudaraster.}

    \label{tab:cudaraster}
\end{table}

We justify this gap in performance by identifying the difference between the design motivations of the two implementations. Cudaraster is extremely specialized to NVIDIA's GPU architecture and hand-optimized to achieve maximum performance for triangle rasterization on that architecture. It benefits from several architecture-specific features that are difficult to generalize into a higher-level abstraction, including its use of core-local shared memory to stage data before and after fetching from off-chip memory, and use of texture caches to accelerate read-only memory access.

While a \piko{} implementation also prioritizes performance and is built with knowledge of the memory hierarchy, we have designed it with equal importance to programmability, flexibility, and portability. We believe that achieving performance within 3--6$\times$ of cudaraster while maintaining these goals demonstrates \piko{}'s ability to realize efficient pipeline implementations.

\subsubsection{Reyes Micropolygon Renderer}

\begin{figure}
\begin{center}
\includegraphics[width=0.48\columnwidth]
    {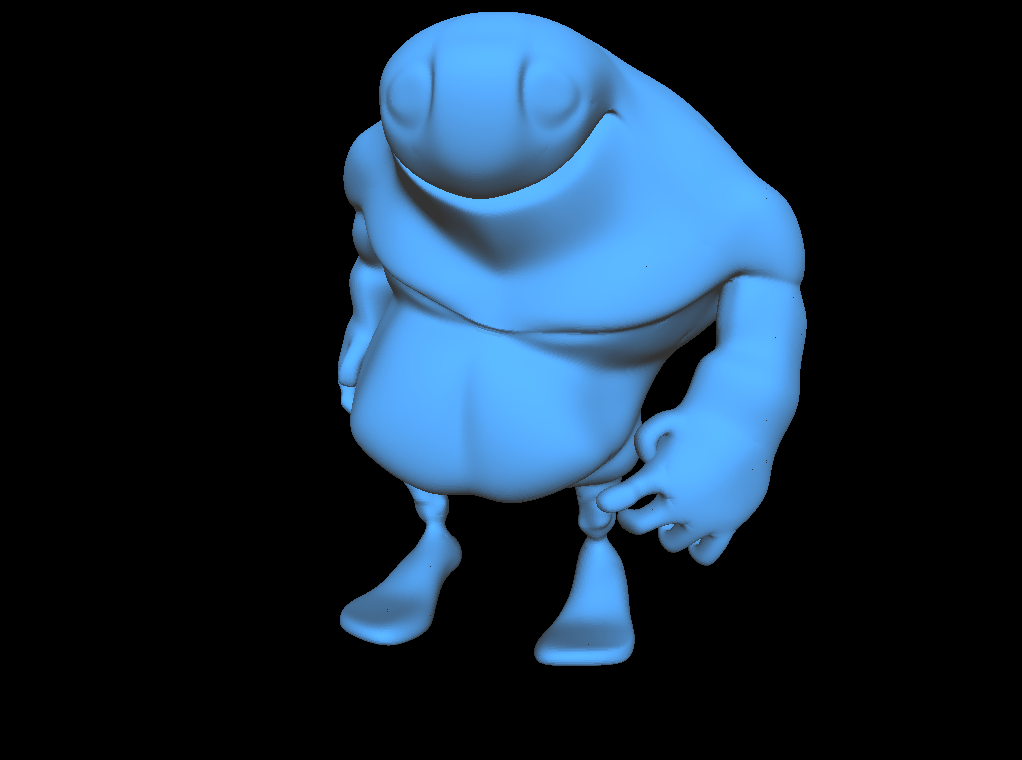}
\includegraphics[width=0.48\columnwidth]
    {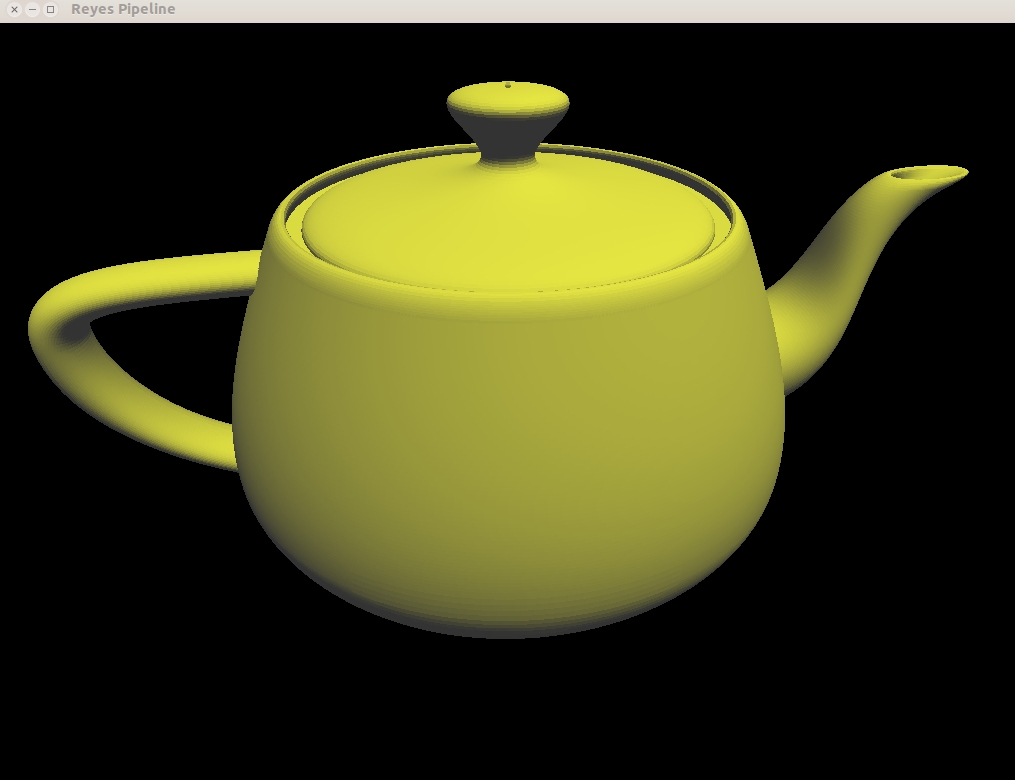}

\caption{Test scenes from our Reyes pipeline generated by \pikoc{}.  The left
scene, Bigguy, renders at 127.6 ms. The right scene, Teapot, renders at 85.1
ms.}

\label{fig:reyespics}

\end{center}
\end{figure}

\figref{reyespics} shows the performance of our Reyes renderer with two
scenes: Bigguy and Teapot. We compared the performance of our Reyes renderer to
Micropolis~\cite{Weber:2015:PRA}, a recently-published Reyes implementation with
a similarly-structured split loop to ours.

We compared the performance of our \stage{Split} stage, written using \piko, to
Micropolis's ``Breadth'' split stage. On the Teapot scene, we achieved 9.44
million patches per second (Mp/s), vs.\ Micropolis's 12.92 Mp/s; while a direct
comparison of GPUs is difficult, Micropolis ran on an AMD GPU with 1.13x more
peak compute and 1.11x more peak bandwidth than ours.
Our implementation is 1.37x
slower, but Micropolis is a complex, heavily-optimized and -tuned implementation
targeted to a single GPU, whereas ours is considerably simpler, easier to write,
understand, and modify, and more portable. For many scenarios, these concerns
may outweigh modest performance differences.

\section{Conclusion}
\label{sec:conclusion}

Programmable graphics pipelines offer a new opportunity for existing and novel
rendering ideas to impact next-generation interactive graphics. Our
contribution, the \piko{} framework, targets high performance with
programmability. \piko's main design decisions are the use of binning
to exploit spatial locality as a fundamental building block for
programmable pipelines, and the decomposition of pipeline stages into
\assignbin{}, \schedule{} and \process{} phases to enable high-level
exploration of the optimization alternatives. This helps implement
performance optimizations and enhance programmability and portability.
More broadly, we hope our work contributes to the conversation about
how to \emph{think} about implementing programmable pipelines.

One of the most important challenges in this work
is how a high-level programming system can enable the important
optimizations necessary to generate efficient pipelines. We believe
that emphasizing spatial locality, so prevalent in both
hardware-accelerated graphics and in the fastest programmable
pipelines, is a crucial ingredient in efficient implementations. In
the near future, we hope to investigate two extensions to this work,
non-uniform bin sizes (which may offer better load balance) and
spatial binning in higher dimensions (for applications like voxel
rendering and volume rendering). Certainly special-purpose hardware
takes advantage of binning in the form of hierarchical or tiled
rasterizers, low-precision on-chip depth buffers, texture derivative
computation, among others. However, our current binning
implementations are all in software. Could future GPU programming
models offer more native support for programmable spatial locality and bins?

From the point of view of programmable graphics,
Larrabee~\cite{Seiler:2008:LAM} took a novel approach: it eschewed
special-purpose hardware in favor of pipelines that were programmed at
their core in software. One of the most interesting aspects of
Larrabee's approach to graphics was software scheduling. Previous
generations of GPUs had relied on hardware schedulers to distribute
work to parallel units; the Larrabee architects instead advocated
software control of scheduling. \piko{}'s schedules are largely
statically compiler-generated, avoiding the complex implementations of
recent work on GPU-based software schedulers, but dynamic, efficient
scheduling has clear advantages for dynamic workloads (the reason we
leverage the GPU's scheduling hardware for limited distribution within
\piko{} already). Exploring the benefits and drawbacks of more general
dynamic work scheduling, such as more task-parallel strategies, is
another interesting area of future work.

In this paper, we consider pipelines that are entirely implemented in
software, as our underlying APIs have no direct access to
fixed-function units. As these APIs add this functionality (and this
API design alone is itself an interesting research problem), certainly
this is worth revisiting, for fixed-function units have significant
advantages. A decade ago, hardware designers motivated fixed-function
units (and fixed-function pipelines) with their large computational
performance per area; today, they cite their superior power
efficiency.  Application programmers have also demonstrated remarkable
success at folding, spindling, and mutilating existing pipelines to
achieve complex effects for which those pipelines were never designed.
It is certainly fair to say that fixed-function power efficiency and
programmer creativity are possible reasons why programmable pipelines
may never replace conventional approaches. But thinking about
pipelines from a software perspective offers us the opportunity to
explore a space of alternative rendering pipelines, possibly as
prototypes for future hardware, that would not be considered in a
hardware-focused environment. And it is vital to perform a fair
comparison against the most capable and optimized programmable
pipelines, ones that take advantage of load-balanced parallelism and
data locality, to strike the right balance between hardware and
software. We hope our work is a step in this direction.


\bibliographystyle{acmsiggraph}
\bibliography{temp_refs,bib/all,bib/owens}
\end{document}